\documentclass[11pt,a4paper]{article}
\usepackage{jheppub}
\usepackage[T1]{fontenc}
\usepackage{amssymb}
\usepackage{mathtools}
\usepackage{stackengine}
\usepackage{scalerel}

\usepackage{hyperref}
\usepackage{epsfig}
\usepackage{amsmath,latexsym,amssymb}
\usepackage{graphicx}
\usepackage{braket}
\usepackage{pdfsync}

\usepackage{framed}
\usepackage{mathtools}

\usepackage{jheppub}
\usepackage[T1]{fontenc}
\usepackage{amssymb}
\usepackage{mathtools}
\usepackage{amsmath}
\usepackage{bm}
\usepackage{stackengine}

\usepackage{scalerel}
\usepackage{eufrak}
\usepackage{framed}

\usepackage{mathtools}

\usepackage{pstricks}

\usepackage{amsmath}
\usepackage{bbold}
\usepackage{bm}
\usepackage{latexsym}
\usepackage{braket}
\usepackage{slashed}
\usepackage{graphicx,booktabs,multirow}
\usepackage{eufrak}
\numberwithin{equation}{section}

\usepackage{color}
\usepackage{pstricks}
\usepackage{amssymb}

\makeatletter
\newcommand{\doublewidetilde}[1]{{%
  \mathpalette\double@widetilde{#1}%
}}
\newcommand{\double@widetilde}[2]{%
  \sbox\z@{$\m@th#1\widetilde{#2}$}%
  \ht\z@=.9\ht\z@
  \widetilde{\box\z@}%
}
\makeatother

\usepackage{hyperref}
\usepackage{slashed}
\usepackage{epsfig}
\usepackage{amsmath,latexsym,amssymb}
\usepackage{graphicx}
\usepackage[latin1]{inputenc}
\usepackage{braket}
\usepackage{pdfsync}

\usepackage{framed}

\usepackage{mathtools}

\def\be{\begin{equation}}
\def\ee{\end{equation}}
\def\ba{\begin{eqnarray}}
\def\ea{\end{eqnarray}}

\newcommand{\bz}{\bar{z}}

\newcommand{\bh}{\bar{h}}

\newcommand{\bx}{\bar{x}}

\newcommand{\zbar}{\bar{z}}

\newcommand{\hy}{{}_{2}F_1}

\def\sv{{\rm sv}}
\def\SVM{{\zeta^m_{\rm sv}}}
\def\SV{{\zeta_{\rm sv}}}

\def\Hc{{\cal H}}

\def\mz#1{\zeta_{#1}}

\def\IQ{{\bf Q}}
\def\IN{{\bf N}}
\def\IR{{\bf R}}
\def\IC{{\bf C}}

\def\IZ{{\bf Z}}
\def\cf{{\it cf.\ }}
\def\sv{{\rm sv}}
\def\SVM{{\zeta^m_{\rm sv}}}
\def\SV{{\zeta_{\rm sv}}}

\def\IZ{ {\bf Z}}\def\IC{{\bf C}}\def\IR{ {\bf R}}
\def\IN{ {\bf N}}
\def\IQ{ {\bf Q}}

\def\z{\zeta}

\def\D{\Delta}

\newcommand{\comment}[1]{}

\def\fc#1#2{{\frac{#1}{#2}}}
\newcommand{\req}[1]{(\ref{#1})}

\def\Lc {{\cal L}}

\def\Ic {{\cal I}}

\def\Hc{{\cal H}}

\def\Ic{{\cal I}}

\def\ov{\overline}

\newcommand{\eea}{\end{eqnarray}}
\def\lf{\left}
\def\ri{\right}
\def\z{\zeta}
\def\ra{{\rightarrow}}

\def\ds{\displaystyle}

\def\sv{{\rm sv}}

\author{
Wei Fan${}^{1}$, Angelos Fotopoulos${}^{2,3}$, Stephan Stieberger${}^{4}$,
Tomasz R.\ Taylor${}^{2}$,\, Bin Zhu${}^2$\\[0.5cm] $^1$\it
Department of Physics, School of Science, Jiangsu University of Science and Technology,
Zhenjiang, 212003, China\\[0.2cm]
 $^2${\it Department of Physics \\
  Northeastern University, Boston, MA 02115, USA}\\[0.2cm]
 $^3${\it Department of Sciences\\
Wentworth Institute of Technology, Boston, MA 02115, USA} \\[.2 cm]
$^4${\it Max--Planck--Institut f\"{u}r Physik,	Werner--Heisenberg--Institut, \\80805 M\"unchen, Germany}}

\title{\boldmath Conformal Blocks from Celestial Gluon Amplitudes\\
II: Single-valued Correlators \unboldmath}

\abstract{In a recent paper, here referred to as part I, we considered the celestial four-gluon amplitude with one gluon represented by the shadow transform of the corresponding primary field operator. This correlator is ill-defined because it contains branch points related to the presence of conformal blocks with complex spin. In this work, we adopt a procedure similar to minimal models and construct a single-valued completion of the shadow correlator, in the limit when the shadow is ``soft.'' By following the approach of Dotsenko and Fateev, we obtain an integral representation of such a single-valued correlator. This allows inverting the shadow transform and constructing a single-valued celestial four-gluon amplitude. This amplitude is drastically different from the original Mellin amplitude. It is defined over the entire complex plane and has correct crossing symmetry, OPE and bootstrap properties. It agrees with all known OPEs of celestial gluon operators. The conformal block spectrum consists of primary fields with dimensions $\Delta=m+i \lambda$, with integer $m\geq 1$  and various, but always integer spin, in all group representations contained in the product of two adjoint representations.}


\begin{document}
\maketitle
\section{Introduction}
In a recent paper \cite{I}, here referred to as part I, we examined the conformal block content of celestial four-gluon amplitudes, that is
Mellin transforms of standard amplitudes with respect to the energies of external particles
\cite{Strominger1706}. Mellin transforms convert the plane wave basis of external wave functions into the ``boost basis'' characterized by conformal dimensions, the variables dual to energies
\cite{Pasterski1705}.\footnote{For recent reviews, see Refs.\cite{Raclariu:2021zjz,Pasterski:2021rjz}} After the  amplitudes are converted to the boost basis, they transform under $SL(2,C)$ Lorentz transformations as two-dimensional CFT correlators of primary fields. The goal of celestial holography \cite{Strominger:2017zoo} is to identify celestial conformal field theory (CCFT) underlying such correlators, and conformal block decomposition is an important step in this direction.

When viewed as CFT correlators, celestial amplitudes exhibit some unusual properties.
Three-gluon amplitudes vanish on-shell due to kinematic constraints \cite{Taylor:2017sph}. On the other hand, three-point correlation functions are the basic building blocks of CFT and should be determined by the operator product expansions (OPE) of primary fields. Four-point CFT correlators depend on one complex variable, $z$, the cross-ratio of four points. In celestial amplitudes, $z$ is a kinematic variable determined by two angles which determine the direction of motion of one particle with respect to the plane spanned by the momenta of the other (three) particles. The scattering processes are planar, as reflected by the constraint $\Im(z)=0$ implied by momentum conservation. This means that four-point correlators vanish everywhere on the complex plane of $z$, with the exception of the real axis. In fact, four-gluon celestial amplitudes are distribution-valued as $\delta(z-\bz)$, which is highly unusual.

Having such a distribution-valued four-point function, one has to make a choice whether to perform conformal block decomposition ``as it is'' \cite{Lam:2017ofc,Nandan:2019jas,Law:2020xcf,Atanasov:2021cje} or to consider some extensions to the entire complex plane of $z$. In part I \cite{I}, we chose the latter route and replaced one conformal (boost) external wave function by its shadow transform. One reason for studying such ``shadowed'' celestial amplitudes is a possible connection between adjoint operators and shadow fields, which could relate four-dimensional (asymptotic) shadow wave functions to {\it in\/} and {\it out\/} states of two-dimensional CFT \cite{spas,I,Crawley:2021ivb,Sharma:2021gcz}. We were able to identify conformal blocks of primary fields with dimensions $\Delta=m+i\lambda$, where $m\ge 2$ is an integer, with various integer spins and in various gauge group representations. There were no blocks with $m=1$ that could describe primary field operators associated with massless gluons.

Four-point CFT correlators obtained from shadowed gluon amplitudes are not completely satisfactory though. At the leading OPE order, in the limit of the coinciding insertion points of two unshadowed gluon operators, the gluons fuse into a single gluon \cite{Fan1903,Strominger1910} and the four-gluon amplitude degenerates into a three-gluon amplitude. As mentioned before, the latter one is zero, therefore, as expected, one does not find any trace of the leading order OPE expansion term in the conformal block decomposition of four-gluon amplitude. In other words, the block with $\Delta=1+i\lambda$ is missing. This is mathematically consistent but unsatisfactory because such four-point correlators seem to be disconnected from well-known OPE coefficients.
Another salient feature of shadowed amplitudes is that in addition to integer spin, they contain conformal blocks with continuous complex spin. This is a manifestation of a serious problem that can be seen even prior to conformal block decomposition: the four-point correlator derived in part I has nontrivial monodromies (branching points), therefore it is {\em not\/} a single-valued function of the cross-ratio $z$. In this work, we resolve both problems simultaneously by constructing a single-valued correlation function that reproduces all known OPEs \cite{Fan1903,Strominger1910} of gluon operators. In the end, we invert the shadow transform and obtain a ``single-valued'' celestial amplitude.  It has a form drastically different from the original Mellin celestial amplitude. It is defined over the entire complex plane and has correct crossing symmetry, OPE and bootstrap properties.

The paper is organized as follows. In Section 2, we start from the four-point correlator derived in part I and discuss its monodromy properties in the limit of ``soft'' shadow with $\Delta=1$. We show that it can be made single-valued by adding just one function, in a similar way as in  minimal models. We perform conformal block decomposition of the resultant correlator in Section 3. In Section 4, we discuss crossing symmetry and bootstrap conditions. We  find complete agreement with all known OPEs \cite{Fan1903,Strominger1910}. We also display some typical contributions to two-gluon OPEs from blocks with $\Delta= m+i\lambda$, $m>1$. In Section 5, we discuss two integral representations of the single-valued correlator: one in terms of integrals over the complex plane and one in terms of  single-valued projections of  line integrals. In Section 6, we invert the shadow transform and obtain the four-gluon, single-valued amplitude.

Similar to part I,  this paper is heavily loaded with hypergeometric functions and we use their properties in many places. Whenever possible, we list relevant identities, but unfortunately some omissions are unavoidable, therefore we refer the reader to chapter 15 of Ref.\cite{abram} for a complete list of hypergeometric identities used in this paper.

\section{Single-valued completion in the soft shadow limit}
The conformal correlators discussed in part I were obtained from the celestial four-gluon amplitude \cite{Strominger1706} in MHV helicity configuration $(-,-,+,+)$, with gauge indices $(a_1,a_2,a_3,a_4)$ and conformal dimensions
$(\Delta_1,\Delta_2,\Delta_3,\Delta_4),~\Delta_i=1+i\lambda_i$ with $\sum_{i=1}^4\lambda_i=0$.
The shadow transform was applied to the first gluon, changing its dimension and helicity:
$\Delta_1\to\tilde\Delta_1=1-i\lambda_1$, $-\to +$. As pointed out in part I, such correlators acquire a simple form in the ``soft'' shadow limit of $\lambda_1=0$. Here, we consider all correlators in this limit, leaving the general case for future work.

Our starting point is the correlator written in Eq.(I.4.14) of part I,\footnote{We add prefix I to equation numbers from part I.} which can be expressed as
\begin{align}
G_{34}^{21}(x,\bar{x})_{s,\lambda_1=0}
&~=~f^{a_1a_2b}f^{a_3a_4b}S_1(x)\bar{I}_1(\bar{x})+f^{a_1a_3b}f^{a_2a_4b}\tilde{S}_1(x)\bar{I}_1(\bar{x}),
\label{rep}\end{align}
where the common (antiholomorphic) factor is given by
\be \bar{I}_1(\bar{x})=(1-\bar{x})^{-1+i\lambda_4}\bar{x}^{1+i\lambda_2}
=\bx^{1+i\lambda_2} {}_2F_1\left({2+i\lambda_2, 1-i\lambda_4\atop 2+i\lambda_2} ;\bx\right)\label{i1fac}\ee
and
\begin{align}
{S}_1(x)&~=~(1-x)^{1+i\lambda_4}x^{-1+i\lambda_2} {}_2F_1\left({2,1-i\lambda_3\atop 1+i\lambda_2};x\right)B(1-i\lambda_3,-i\lambda_4)\ ,\label{s1def}\\
\tilde{S}_1(x)&~=~-(1-x)^{1+i\lambda_4}x^{-1+i\lambda_2} {}_2F_1\left({2,-i\lambda_3\atop i\lambda_2};\,x\right)B(-i\lambda_3,-i\lambda_4)\ .\label{s1tdef}
\end{align}
Note that in the cross-ratio
\be x=\frac{z_{12}z_{34}}{z_{13}z_{24}}\ ,\label{xdef}\ee
$z_1$ is the position of the shadowed gluon operator.\footnote{In a slight change of notation from part I, the shadow point is now denoted by $z_1$ instead of $z_1'$. We will revert to the original notation in Section 6, where we invert the shadow transform.} We want to discuss the monodromy properties of the correlator (\ref{rep}). We begin with the term associated with the first group factor, {\em i.e}.\ on $S_1(x)\bar{I}_1(\bar{x})$. It is single-valued, $\sim x^{-2}|x|^{2+2i\lambda_2}$ near $x=0$. In order to examine monodromy at $x=1$, we analytically continue the hypergeometric function to $x\approx 1$:
\begin{align}
S_1(x)
&=(1-x)^{1+i\lambda_4}x^{i\lambda_2-1}B(1-i\lambda_3,-2-i\lambda_4) \,_2F_1\left({2,\,1-i\lambda_3\atop 3+i\lambda_4};\, 1-x\right)\nonumber\\[1mm]
& \qquad +~\frac{\Gamma(-i\lambda_4)\Gamma(2+i\lambda_4)}{x(1-x)} \, _2F_1\left({-1,\,i\lambda_3, \atop -i\lambda_4-1}; \, 1-x\right)\ .\label{s1var}
\end{align}
Note that the second term is a rational function:
\be
\frac{\Gamma(-i\lambda_4)\Gamma(2+i\lambda_4)}{x(1-x)} \,_2F_1\left({-1,\,i\lambda_3, \atop -i\lambda_4-1}; \, 1-x\right) =\Gamma(-i\lambda_4)\Gamma(1+i\lambda_4)\frac{1-i\lambda_2-i\lambda_3 \, x}{x(1-x)} \, ,\label{eq:S1I1bad}
\ee
therefore the respective contribution  to $S_1(x)\bar{I}_1(\bar{x})$
is \be
\Gamma(-i\lambda_4)\Gamma(1+i\lambda_4)\frac{1-i\lambda_2-i\lambda_3 \, x}{x(1-x)}(1-\bar{x})^{-1+i\lambda_4}\bar{x}^{1+i\lambda_2},\label{exterm}\ee
which behaves as
$\sim |1-x|^{-2} (1-\bar{x})^{i\lambda_4}$ near $x=1$ and has a nontrivial monodromy. Note that the contribution of the first term of Eq.(\ref{s1var}) to $S_1(x)\bar{I}_1(\bar{x})$ is single-valued in this neighborhood. Hence, in order to construct a single-valued correlator, we need to supplement (\ref{exterm}) with some other contributions. At first glance, this looks like a vague, hopeless task, however an important hint comes from minimal models where a similar problem appears when combining  insertions of ``charge-screening'' operators \cite{DiF,dots}. In a minimal model with a Verma module degenerating at level 2, such contributions involve  a single conformal block and its shadow block \cite{Osborn:2012vt}.

In part I, we pointed out [{\em c.f}.\ Eqs.(I.4.17)-(I.4.19)] that the antiholomorphic part $\bar{I}_1(\bar{x})$ describes a single chiral conformal block with $\bar h=1+\frac{i\lambda_2}{2}$. Its shadow block with  weight $1-\bar h=-\frac{i\lambda_2}{2}$ is given by
\be
\bar{I}_2(\bar{x}) = \left.\bar{x}^{1-\bar{h}-\bar{h}_3-\bar{h}_4} \,_2F_1\left({ 1-\bar{h}-\bar{h}_{12},\, 1-\bar{h}+\bar{h}_{34}\atop 2-2\bar{h}};\bar{x}\right)\right|_{\bar{h}=1+\frac{i\lambda_2}{2}}= {} _2F_1\left({1,i\lambda_3\atop -i\lambda_2};\bar{x}\right)\ . \label{eq:I2bar}
\ee
When analytically continued to $x\approx 1$, it becomes
\be
\bar{I}_2(\bar{x})=\frac{1+i\lambda_2}{1-i\lambda_4}\, _2F_1\left({1,i\lambda_3\atop 2-i\lambda_4}; 1-\bar{x}\right)+ \frac{\Gamma(-i\lambda_2)\Gamma(1-i\lambda_4)}{\Gamma(i\lambda_3)} \bar{x}^{1+i\lambda_2}(1-\bar{x})^{-1+i\lambda_4}.
\ee
We see that the second term shares the $\bar{I}_1(\bar{x})$ factor (\ref{i1fac}) with the unwanted term (\ref{exterm}).
Now it becomes clear how to cancel this term and construct a single-valued combination. It is
\be S_1(x)\bar{I}_1(\bar{x}) +S_2(x)\bar{I}_2(\bar{x})\label{svcom}\ee
with
\be
S_2(x) = B(i\lambda_3,i\lambda_4)\frac{1-i\lambda_2-i\lambda_3 \, x}{x(1-x)}
=B(i\lambda_3,i\lambda_4) \frac{1-i\lambda_2}{x(1-x)}\,  {}_2F_1\left({-1, i\lambda_3\atop 1-i\lambda_2}; x\right)
\, .\label{s2def}
\ee
Indeed, the term with a monodromy at $x=1$ is absent in this combination. Finally, by using well-known properties of hypergeometric functions, we can analytically continue (\ref{svcom}) from $x$ to $1/x$  to verify that, as expected, there is no monodromy at infinity. We conclude that this combination is single-valued on the entire complex plane.

We can repeat the same procedure for the coefficient of the second group factor in Eq.(\ref{rep}),
to find the following single-valued combination:
\be \tilde S_1(x)\bar{I}_1(\bar{x}) +\tilde S_2(x)\bar{I}_2(\bar{x})\label{svcom2},
\ee
where
\be
\tilde S_2(x)=-B(i\lambda_3,i\lambda_4)\frac{2-i\lambda_2-(1+i\lambda_3)x}{1-x} =B(i\lambda_3,i\lambda_4)
\frac{i\lambda_2-2}{1-x}\, _2F_1\left({-1,\, 1+i\lambda_3\atop 2-i\lambda_2}; \, x\right)\label{s2tdef}
\ee
As a result, we obtain the following single-valued correlator
\begin{align}
G_{34}^{21}(x,\bar{x})_{SV}
&~=~f^{a_1a_2b}f^{a_3a_4b}[S_1(x)\bar{I}_1(\bar{x})+S_2(x)\bar{I}_2(\bar{x})]\nonumber \\[1mm]
&\qquad +f^{a_1a_3b}f^{a_2a_4b}[\tilde{S}_1(x)\bar{I}_1(\bar{x})+\tilde{S}_2(x)\bar{I}_2(\bar{x})]\ .\label{repf}\end{align}
It is remarkable that a single-valued correlation function can be constructed by adding just one term for each group factor. But this is just the beginning of forthcoming miracles...
\section{Conformal block decomposition}
The correlator (\ref{rep}) was obtained in part I by taking a shadow transform of the celestial amplitude describing incoming particles 1 and 2 scattering into outgoing particles 3 and 4. In such a process,  Mandelstam variables are constrained by $s>0$, $t<0,~u<0$ and the integration region of the shadow transform is subject to the respective constraint on the cross-ratio of celestial coordinates. Thus, the single-valued function (\ref{repf}) should be identified with the following four-gluon correlator:\footnote{As mentioned before, all correlators are considered in the limit of $\lambda_1=0$. The shadow gluon field has $(h_1,\bar{h}_1)=(1,0)$.
}
\be
G_{34}^{21}(x,\bar{x})_{SV} =  \lim_{z_{1}, \bar{z}_{1}\rightarrow \infty} z_{1}^2
\Big\langle\tilde\phi_{\tilde\D_1=1,+}^{a_1,-\epsilon}(z_{1},\zbar_{1})\,
\phi_{\D_2,-}^{a_2,-\epsilon}(1,1)
\,\phi_{\D_3,+}^{a_3,+\epsilon}
(x,\bar x)
\,\phi_{\D_4,+}^{a_4,+\epsilon}(0,0)\Big\rangle\ ,
\label{gcorf}
\ee
where we used the same notation as in part I. In addition, we introduced the indices $\pm\epsilon$
to distinguish between outgoing and incoming particles, respectively \cite{Strominger1910}.
We should stress once again that this correlator describes four-dimensional ``$s$-channel'' gluon collisions $12\to 34$ and this will remain unchanged throughout the present paper.\footnote{While in part I, we also considered four-dimensional $u$- and $t$-channels.} On the other hand,
the choice of the shadow point $z_1$ is arbitrary, therefore the cross-ratio $x$ of Eq.(\ref{xdef}) in {\em not\/}
constrained by kinematics -- it can take any value on the complex plane. Thus, we are free to explore the regions of $x\approx 0$ [$(12\rightleftharpoons 34)_{{\mathfrak{2}}}$ CFT channel],
$x\approx 1$ [$(14\rightleftharpoons 32)_{{\mathfrak{2}}}$ CFT channel]
and $x\approx \infty$ [$(13\rightleftharpoons 42)_{{\mathfrak{2}}}$ CFT channel].
To that end,  we will expand the correlator (\ref{gcorf}) in powers of $x$ , $1-x$ and $1/x$, and decompose it into the corresponding conformal blocks.
\subsection{$\mathbf{(12\rightleftharpoons 34)_\mathfrak{2}}$ blocks}
In this channel, a conformal block of a primary field with chiral weights $(h,\bar h)$ has the form \cite{Osborn:2012vt} (see also (I.4.17)):
\be K_{34}^{21}[h,\bh]=\bx^{\bh-\bh_3-\bh_4}\hy\left({\bh-\bh_{12},\bh+\bh_{34}\atop 2\bh};\bx\right)x^{h-h_3-h_4}\hy\left({h-h_{12},h+h_{34}\atop 2h};x\right)\ ,\label{blcs}\ee
where $h_{ij}=h_i-h_j$. In our case
\begin{align}\nonumber&
h_{12}= 1-\textstyle \frac{i\lambda_2}{2}\ ,~  \bh_{12}=-1-\textstyle\frac{i\lambda_2}{2},\qquad\qquad
h_{34}=\textstyle\frac{i\lambda_3}{2}-\textstyle\frac{i\lambda_4}{2}
\ ,~  \bh_{34}=\textstyle\frac{i\lambda_3}{2}-\textstyle\frac{i\lambda_4}{2}\ ,\\[1mm] &
h_3+h_4=2+\textstyle\frac{i\lambda_3}{2}+\textstyle\frac{i\lambda_4}{2}=2-\textstyle
\frac{i\lambda_2}{2}\ , ~~~~~~~~\bh_3+\bh_4=\textstyle\frac{i\lambda_3}{2}+\textstyle\frac{i\lambda_4}{2}=-
\textstyle\frac{i\lambda_2}{2}\ ,
\end{align}
where we used $\lambda_2+\lambda_3+\lambda_4=0$.

In part I, we performed conformal block decomposition of the original correlator (\ref{rep}) by using Gau{\ss} recursion relations and basic properties of hypergeometric functions.
In particular, we showed that its anti-holomorphic part is associated with a single weight: in the $s$-channel $\bar{I}_1$ represents a chiral block with $\bar{h}=1+i\lambda_2/2$.
A similar procedure can be applied to the single-valued completion (\ref{repf}). We have already identified $\bar{I}_2$ as the shadow block with $\bar{h}=-i\lambda_2/2$, therefore it remains to rewrite $S_2$ and $\tilde S_2$ in terms of the hypergeometric functions present in Eq.(\ref{blcs}). At the end, we find:
\begin{align}
G_{34}^{21}(x,\bx)_{SV} =& \sum_{m=1}^{\infty}
(a_m\, f^{a_1a_2b}f^{a_3a_4b}+\tilde a_m\, f^{a_1a_3b}f^{a_2a_4b})
K_{34}^{21}\Big[m+\frac{i\lambda_2}{2} ,1+\frac{i\lambda_2}{2}\Big](x,\bx) \nonumber\\
&+\sum_{m=1}^{\infty}(b_m\, f^{a_1a_2b}f^{a_3a_4b}+\tilde b_m\, f^{a_1a_3b}f^{a_2a_4b})
K_{34}^{21}\Big[m-\frac{i\lambda_2}{2} , -\frac{i\lambda_2}{2}\Big](x,\bx)
,\label{gss1}
 \end{align}
 with the following coefficients:
\begin{align}\label{coeff1}
a_m& =\frac{m!\,\Gamma(-i\lambda_3+m)\Gamma(-i\lambda_4)}{\Gamma(i\lambda_2+2m-1)}\ ,\\[1mm]\label{coeff2}
\tilde a_m&=-a_m+(-1)^m\frac{m!\,\Gamma(-i\lambda_4+m)\Gamma(-i\lambda_3)}{\Gamma(i\lambda_2+2m-1)}\ ,\\[1mm]
b_m &= \frac{\Gamma(i\lambda_3)\Gamma(m+i\lambda_4)\Gamma(1-i\lambda_2+m)}{i\lambda_4 \, \Gamma(-i\lambda_2)\Gamma(2m-i\lambda_2-1)}\, ,  \label{eq:bm} \\[1mm]
\tilde{b}_m &= -b_m -(-1)^m \frac{\Gamma(i\lambda_4)\Gamma(m+i\lambda_3)\Gamma(1-i\lambda_2+m)}{i\lambda_3 \, \Gamma(-i\lambda_2)\Gamma(2m-i\lambda_2-1)}\, .\label{eq:btm}
\end{align}
In this way, we identify two sets of conformal blocks. The original one has $(h,\bar{h}) = (m+\frac{i\lambda_2}{2}, 1+\frac{i\lambda_2}{2})$, with $m\ge 1$. They have dimensions $\Delta = 2+J+i\lambda_2$, where $J\ge 0$ is an integer spin. Its completion comes with $(h,\bar{h}) = (m-\frac{i\lambda_2}{2}, -\frac{i\lambda_2}{2})$, with $m\ge 1$. They have dimensions $\Delta = J-i\lambda_2= J+i\lambda_3+i\lambda_4$, where $J\ge 1$ is an integer spin. Note that the second, new set contains a block with spin $J=1$ and dimension $\Delta=1+i\lambda_3+i\lambda_4$. This is exactly what one expects from the leading term in the OPE of $\phi_{\D_3,+}^{a_3,+\epsilon}(x,\bar x)
\,\phi_{\D_4,+}^{a_4,+\epsilon}(0,0)$. It appears that the single-valued completion of the correlator restores correct OPEs. This will be confirmed in Section 4 where we discuss crossing symmetry and OPEs.
\subsection{$\mathbf{(14\rightleftharpoons 32)_{{\mathfrak{2}}}}$ blocks}
In this channel, a conformal block of a primary field with chiral weights $(h,\bar h)$ has the form:
\begin{align} K_{32}^{41}[h,\bh](1-x,1-\bx)~=~ & (1-x)^{h-h_3-h_2}\hy\left({h-h_{14},h+h_{32}\atop 2h}; 1-x\right)\nonumber\\[1mm] &\times(1-\bx)^{\bh-\bh_3-\bh_2}\hy\left({\bh-\bh_{14},\bh+\bh_{32}\atop 2\bh};1-\bx\right) ,\label{blsu}\end{align}
where
\begin{align}\nonumber&
h_{14}= -\textstyle \frac{i\lambda_4}{2}\ , ~\bh_{14}=-\textstyle\frac{i\lambda_4}{2}\ ,\qquad\qquad
h_{32}=1+\textstyle\frac{i\lambda_3}{2}-\frac{i\lambda_2}{2}\ ,~
\bh_{32}=-1+\textstyle\frac{i\lambda_3}{2}-\frac{i\lambda_2}{2}\ ,\\[1mm] &
h_3+h_2=1+\textstyle\frac{i\lambda_3}{2}+\textstyle\frac{i\lambda_2}{2}=1-\textstyle\frac{i\lambda_4}{2}\ , ~~~~~~~~~\bh_3+\bh_2=1+\textstyle\frac{i\lambda_3}{2}+\textstyle\frac{i\lambda_2}{2}=
1-\textstyle\frac{i\lambda_4}{2}\ .
\end{align}
In order to decompose the correlator (\ref{repf}) into the blocks (\ref{blsu}), we need to rewrite it as
\begin{align}
G_{34}^{21}(x,\bx)_{SV}& ~=~G_{32}^{41}(1-x,1-\bx)_{SV} \nonumber\\[1mm] &
~=~f^{a_1a_2b}f^{a_3a_4b}G_u(1-x,1-\bx)+f^{a_1a_3b}f^{a_2a_4b}\tilde G_u(1-x,1-\bx)\ , \label{gsu1}
\end{align}
where $G_u$ and $\tilde G_u$ are obtained from the single-valued combinations (\ref{svcom}) and (\ref{svcom2}), respectively, by analytic continuation to the region of $x\approx 1$, as described in the previous section. We find
\begin{align}
G_u&~=\, B(1-i\lambda_3,-2-i\lambda_4) x^{i\lambda_2-1}(1-x)^{1+i\lambda_4} \, _2F_1\Bigg({2,1-i\lambda_3\atop 3+i\lambda_4}; 1-x\Bigg) \bar{x}^{1+i\lambda_2}(1-\bar{x})^{-1+i\lambda_4} \nonumber\\
&+B(i\lambda_4,i\lambda_3)\frac{(1+i\lambda_4)(1+i\lambda_2)}{1-i\lambda_4}\frac{x^{i\lambda_2-1}}{(1-x)}\, _2F_1\Bigg({-i\lambda_4, i\lambda_2-1 \atop -i\lambda_4-1}; 1-x\Bigg)\, _2F_1\Bigg({i\lambda_3, 1 \atop 2-i\lambda_4}; 1-\bar{x}\Bigg),
\end{align}
and
\begin{align}
\tilde{G}_u&~=\,
B(-i\lambda_3,-2-i\lambda_4)x^{i\lambda_2-1}(1-x)^{1+i\lambda_4}\, _2F_1\Bigg({2,-i\lambda_3 \atop 3+i\lambda_4}; 1-x\Bigg)\bar{x}^{1+i\lambda_2}(1-\bar{x})^{-1+i\lambda_4} \nonumber\\
&- B(i\lambda_3,i\lambda_4)\frac{(1+i\lambda_2)(1+i\lambda_4)}{1-i\lambda_4}\frac{x^{i\lambda_2-1}}{1-x} \, _2F_1\Bigg({-i\lambda_4,-2+i\lambda_2\atop -i\lambda_4-1}; 1-x\Bigg)\, _2F_1\Bigg({i\lambda_3, 1 \atop 2-i\lambda_4}; 1-\bar{x}\Bigg) \, .\end{align}
where in addition to the analytic continuation formulas, we used
\begin{equation}\label{hy1}
\hy\left({a,b\atop c};1-x\right) = x^{c-a-b}\hy\left({c-a,c-b\atop c};1-x\right).\ee
{}From this point, we can proceed with conformal block decomposition in the same way as in the ${(12\rightleftharpoons 34)_{{\mathfrak{2}}}}$ channel. At the end, we obtain
\begin{align}
&G_{34}^{21}(x,\bx)_{SV} =G_{32}^{41}(1-x,1-\bx)_{SV} \nonumber\\[1mm]
&=\sum_{m=1}^{\infty}
(c_m\, f^{a_1a_2b}f^{a_3a_4b}+\tilde c_m\, f^{a_1a_3b}f^{a_2a_4b})
K_{32}^{41}\Big[m+1+\frac{i\lambda_4}{2} ,\frac{i\lambda_4}{2}\Big](1-x,1-\bx) \nonumber\\
&~~+\sum_{m=1}^{\infty}(d_m\, f^{a_1a_2b}f^{a_3a_4b}+\tilde d_m\, f^{a_1a_3b}f^{a_2a_4b})
K_{32}^{41}\Big[m-1-\frac{i\lambda_4}{2} , 1-\frac{i\lambda_4}{2}\Big](1-x,1-\bx) \, , \label{gsu1}
\end{align}
where
\begin{align}
c_m &= \frac{m!\,\Gamma(-2-i\lambda_4)\Gamma(3+i\lambda_4)\Gamma(-i\lambda_3+m)}{\Gamma(-1+i\lambda_2)\Gamma(1+i\lambda_4+2m)}\, ,\\[1mm]
d_m &= B(i\lambda_3,i\lambda_4)\,\frac{1+i\lambda_2}{i\lambda_4-1}\,\frac{\Gamma(-i\lambda_4+m-1)\Gamma(i\lambda_2-2+m)}{\Gamma(i\lambda_2-1)\Gamma(-i\lambda_4-3+2m)}\, ,\label{eq:dm}\\[1mm]
\tilde{c}_m &= (-1)^{m}\,\frac{m!\, \Gamma(-i\lambda_3)\Gamma(3+i\lambda_4)\Gamma(-2-i\lambda_4)}{\Gamma(-2+i\lambda_2)\Gamma(3-i\lambda_2)}\frac{\,\Gamma(2-i\lambda_2+m)}{\Gamma(1+i\lambda_4+2m)}\, , \\[1mm]
\tilde{d}_m &= (-1)^{m}\,B(i\lambda_3,i\lambda_4)\,\frac{1+i\lambda_2}{i\lambda_4-1}\, \frac{\Gamma(-i\lambda_4+m-1)\Gamma(i\lambda_3+m)}{\Gamma(1+i\lambda_3)\Gamma(-i\lambda_4-3+2m)} \, . \label{eq:dtm}
\end{align}
Here again, we find two sets of conformal blocks. The first set comes with $(h,\bar{h}) = (m+1+\frac{i\lambda_4}{2}, \frac{i\lambda_4}{2})$, with $m\ge 1$. They have dimensions $\Delta = J+i\lambda_4$, where $J\ge 2$ is an integer spin. The second set has $(h,\bar{h}) = (m-1-\frac{i\lambda_4}{2}, 1-\frac{i\lambda_4}{2})$, with $m\ge 1$. They have dimensions $\Delta = J+2-i\lambda_4= J+2+i\lambda_2+i\lambda_3$, where $J\ge -1$ is an integer spin. The second set contains a block with $\Delta=1+i\lambda_2+i\lambda_3$ and $J=-1$, which represents the gluon operator appearing in the product $\phi_{\D_2,-}^{a_2,-\epsilon}(1,1)
\,\phi_{\D_3,+}^{a_3,+\epsilon}(x,\bar x)$  at the leading order of OPE of as $x\to 1$. Recall that in part I, we also found blocks with continuous, complex spin. Due to the presence of such exotic states, we called this two-dimensional channel ``incompatible'' with the four-dimensional $s$-channel. Now we see that the single-valued completion eliminates such exotic states. Actually, all channels are perfectly compatible.

\subsection{$\mathbf{(13\rightleftharpoons 42)_{{\mathfrak{2}}}}$ blocks}
In this channel, a conformal block of a primary field with chiral weights $(h,\bar h)$ has the form:
\begin{align} K_{31}^{24}[h,\bh]\Big(\frac{1}{x},\frac{1}{\bx}\Big)=&\Big(\frac{1}{x}
\Big)^{h-h_3-h_{1}}\hy\left({h-h_{42},h+h_{31}\atop 2h}; \frac{1}{x}\right)\nonumber\\
&~~\times\Big(\frac{1}{\bx}\Big)^{\bh-\bh_3-\bh_{1}}\hy\left({\bh-\bh_{42},\bh+\bh_{31}\atop 2\bh};\frac{1}{\bx}\right)\ ,\label{blst}\end{align}
where
\begin{align}
&h_{42} = \textstyle 1+\frac{i\lambda_4}{2}-\frac{i\lambda_2}{2}\, ,  ~\bh_{42} = \textstyle -1+\frac{i\lambda_4}{2}-\frac{i\lambda_2}{2}\, ,\qquad
h_{31}  = \textstyle \frac{i\lambda_3}{2}\, ,~ \bh_{31}  = \textstyle \frac{i\lambda_3}{2} \nonumber\\[1mm]
&h_3+h_{1} = \textstyle 2+\frac{i\lambda_3}{2},\qquad\qquad\qquad\qquad~~~~~~~~~\qquad \bh_3+\bh_{1} = \textstyle \frac{i\lambda_3}{2} .
\end{align}
In order to decompose the correlator (\ref{repf}) into the blocks (\ref{blst}), we need to rewrite it as
\begin{align}&x^{2h_3}\bar{x}^{2\bh_3}
G_{34}^{21}(x,\bx)_{SV} ~=~G_{24}^{13}\Big(\frac{1}{x},\frac{1}{\bx}\Big)_{SV} \nonumber\\[1mm] &
~~~~=~f^{a_1a_2b}f^{a_3a_4b}G_t\Big(\frac{1}{x},\frac{1}{\bx}\Big)+f^{a_1a_3b}f^{a_2a_4b}\tilde G_t\Big(\frac{1}{x},\frac{1}{\bx}\Big)\ ,
\end{align}
where the factor $x^{2h_3}\bar{x}^{2\bh_3}$, with $(h_3,\bh_3)=(1+\frac{i\lambda_3}{2},\frac{i\lambda_3}{2})$, originates from the conformal transformation of the respective operator when the coordinate transforms from $x$ to $1/x$ \cite{DiF}. The functions
$G_t$ and $\tilde G_t$ are obtained from the single-valued combinations (\ref{svcom}) and (\ref{svcom2}), respectively, by analytic continuation to the region of $x\approx \infty$, using the well-known relations between the hypergeometric functions at $x$ and $1/x$.
In this way, we obtain:
\begin{align}
G_t&~=~ B(-i\lambda_4,-1-i\lambda_3) \Big(1-\frac{1}{x}\Big)^{1+i\lambda_4}\, _2F_1\left({2,2-i\lambda_2\atop 2+i\lambda_3}; \frac{1}{x}\right) \Big( 1-\frac{1}{\bx}\Big)^{-1+i\lambda_4}\nonumber\\[2mm]
&+\frac{(1+i\lambda_2)i\lambda_3}{1-i\lambda_3}B(i\lambda_3,i\lambda_4)\frac{x^{2+i\lambda_3}}{1-x}\, _2F_1\left({-1,i\lambda_2-1\atop -i\lambda_3}; \frac{1}{x}\right)\bx^{i\lambda_3-1} \, _2F_1\left({1,2+i\lambda_2\atop 2-i\lambda_3}; \frac{1}{\bx}\right) \, ,
\end{align}
\begin{align}
\tilde{G}_t&~=~ -B(-i\lambda_4,-2-i\lambda_3)\Big( 1-\frac{1}{x}\Big)^{1+i\lambda_4}\, _2F_1\left({2,3-i\lambda_2\atop 3+i\lambda_3}; \frac{1}{x}\right) \Big( 1-\frac{1}{\bx}\Big)^{-1+i\lambda_4}   \nonumber\\[2mm]
&-\frac{(1+i\lambda_2)(1+i\lambda_3)}{1-i\lambda_3}B(i\lambda_3,i\lambda_4)\frac{x^{3+i\lambda_3}}{1-x}\, _2F_1\left({-1,i\lambda_2-2\atop -i\lambda_3-1}; \frac{1}{x}\right) \bx^{i\lambda_3-1} \, _2F_1\left({1,2+i\lambda_2\atop 2-i\lambda_3}; \frac{1}{\bx}\right) \, .
\end{align}
After repeating the same steps as in other channels, we obtain the following conformal block decomposition:
\begin{align}
& x^{2h_3}\bx^{2\bh_3} G^{21}_{34}(x,\bx)_{SV}=G_{31}^{24}\Big(\frac{1}{x},\frac{1}{\bx}\Big)_{SV}\nonumber\\
&=\sum_{m=1}^{\infty}
(e_m\, f^{a_1a_2b}f^{a_3a_4b}+\tilde e_m\, f^{a_1a_3b}f^{a_2a_4b})
K_{31}^{24}\Big[m+1+\frac{i\lambda_3}{2} ,\frac{i\lambda_3}{2}\Big]\Big(\frac{1}{x},\frac{1}{\bx}\Big) \nonumber\\
&~~+\sum_{m=1}^{\infty}(f_m\, f^{a_1a_2b}f^{a_3a_4b}+\tilde f_m\, f^{a_1a_3b}f^{a_2a_4b})
K_{31}^{24}\Big[m-1-\frac{i\lambda_3}{2} , 1-\frac{i\lambda_3}{2}\Big]\Big(\frac{1}{x},\frac{1}{\bx}\Big) \, , \label{gst1}
\end{align}
where
\begin{align}
\tilde{e}_m &= -B(-i\lambda_4,-1-i\lambda_3)\frac{m!\, \Gamma(2+i\lambda_3)\Gamma(2+m-i\lambda_2)}{\Gamma(1+i\lambda_3+2m)\Gamma(2-i\lambda_2)} \, ,\\
\tilde{f}_m &= -B(i\lambda_3,i\lambda_4)\frac{1+i\lambda_2}{1-i\lambda_3}\frac{\Gamma(-i\lambda_3-1+m)
\Gamma(i\lambda_4+m)}{\Gamma(-3-i\lambda_3+2m)\Gamma(1+i\lambda_4)} \, , \label{eq:ftm}\\
e_m &= -\tilde{e}_m -(-1)^m B(-i\lambda_4,-1-i\lambda_3)\frac{m!\,\Gamma(2+i\lambda_3)\Gamma(m-i\lambda_4)}{
\Gamma(1+i\lambda_3+2m)\Gamma(-i\lambda_4)}\, ,\\
f_m & = -\tilde{f}_m +(-1)^m\,B(i\lambda_3,i\lambda_4)\frac{1+i\lambda_2}{1-i\lambda_3}\frac{\Gamma(-i\lambda_3+m-1)
\Gamma(i\lambda_2+m-2)}{\Gamma(-3-i\lambda_3+2m)\Gamma(i\lambda_2-1)} \, . \label{eq:fm}
\end{align}
Here again, we find two sets of conformal blocks. The first set comes with $(h,\bar{h})
 = (m+1+\frac{i\lambda_3}{2}, \frac{i\lambda_3}{2})$, with $m\ge 1$. They have dimensions $\Delta = J+i\lambda_3$, where $J\ge 2$ is an integer spin. The second set has $(h,\bar{h}) = (m-1-\frac{i\lambda_3}{2}, 1-\frac{i\lambda_3}{2})$, with $m\ge 1$. They have dimensions $\Delta = J+2-i\lambda_3= J+2+i\lambda_2+i\lambda_4$, where $J\ge -1$ is an integer spin. Note that when the indices 3 and 4 are interchanged, the spectrum of $(13\rightleftharpoons 42)_{{\mathfrak{2}}}$ conformal blocks in Eq.(\ref{gst1}) becomes identical to the spectrum of $(14\rightleftharpoons 32)_{{\mathfrak{2}}}$ blocks in Eq.(\ref{gsu1}). As explained in Section 4, this is related to the crossing symmetry of CCFT.
\subsection{Channel decomposition of ${ SU(2)}$ group factors}
The conformal block decomposition accomplished so far allows identifying dimensions and spins of primary operators created by gluon fusion. The information about their group representations is contained in group factors that need to be decomposed in the corresponding channels. {}For a given channel  $(ij\rightleftharpoons kl)_{{\mathfrak{2}}}$, the goal is to rewrite the group factors in the form
$\sum_r\alpha_rC_r^{a_ia_j}C_r^{*a_ka
_l}$, where $C_r^{a_ia_j}=\langle r|a_i,a_j\rangle$ are the Clebsch-Gordan coefficients for the fusion of gluons with group indices $a_i$ and $a_j$ into all possible multiplets $r$ contained in the product of two adjoint representations. The coefficients $\alpha_r$ need to be determined. For a general group, this can be done by using rather advanced techniques, which are beyond the scope of the present paper.  Such a decomposition is very simple, however, in  the  case of $SU(2)$. Then gluons are isospin $I=1$ triplets, therefore $r$ can be $I=0$ singlet, $I=1$ triplet or $I=2$ quintuplet and we can use standard Clebsch-Gordan coefficients.

The $SU(2)$ structure constants $f^{abc}=\epsilon^{abc}$, therefore the relevant group factors are
\be
f^{abx}f^{cdx}=\delta^{ac}\delta^{bd}-\delta^{ad}\delta^{bc}.\label{cfact}\ee
$SU(2)$ Clebsch-Gordan coefficients, which are usually tabulated in the angular momentum basis with the multiplet components  $|I,I_3=M\rangle$. In scattering amplitudes, however, gluons are labeled by $a=1,2,3$ vector indices. Therefore, in order to perform the desired channel decomposition, Clebsch-Gordan coefficients need to be converted from the standard
$|1,M\rangle ~(M=0, \pm 1)$ basis to $|a\rangle~ (a=1,2,3)$. This is done in Appendix A, where we give explicit expressions for
$C_{2M}^{ab}=\langle 2,M|a,b\rangle ~\,(M=0,\pm 1,\pm 2)$, $C_{1M}^{ab}=\langle 1,M|a,b\rangle ~\,(M=0,\pm 1)$ and $C_{00}^{ab}=\langle 0,0|a,b\rangle$.
Note that Clebsch-Gordan coefficients are symmetric in group indices $a,b$ for $I=0,2 $ and antisymmetric for $I=1$. Then, by using Eq.(\ref{cfact}), the group factors can be decomposed in the $(12\rightleftharpoons 34)_{{\mathfrak{2}}}$ channel as
\begin{align}
f^{a_1a_2x}f^{a_3a_4x} &= 2\sum_M C_{1M}^{a_1a_2}C_{1M}^{*\, a_3a_4}\, , \label{eq:f1s}\\
f^{a_1a_3x}f^{a_2a_4x} &= -\sum_M C_{2M}^{a_1a_2}C_{2M}^{*\, a_3a_4}+\sum_M C_{1M}^{a_1a_2}C_{1M}^{*\, a_3a_4}+2C_{00}^{a_1a_2}C_{00}^{* \, a_1a_2} \,  . \label{eq:f2s}
\end{align}
Similarly, in the $(14\rightleftharpoons 32)_{{\mathfrak{2}}}$ channel:
\begin{align}
f^{a_1a_2x}f^{a_3a_4x} &= \sum_M C_{2M}^{a_1a_4}C_{2M}^{*\, a_3a_2}+\sum_M C_{1M}^{a_1a_4}C_{1M}^{*\, a_3a_2}- 2C_{00}^{a_1a_4}C_{00}^{*\, a_3a_2}\, , \label{eq:f1u}\\
f^{a_1a_3x}f^{a_2a_4x} &= \sum_M C_{2M}^{a_1a_4}C_{2M}^{*\, a_3a_2}-\sum_M C_{1M}^{a_1a_4}C_{1M}^{*\, a_3a_2}- 2C_{00}^{a_1a_4}C_{00}^{*\, a_3a_2}\,  \label{eq:f2u}
\end{align}
and in $(13\rightleftharpoons 42)_{{\mathfrak{2}}}$:
\begin{align}
f^{a_1a_2x}f^{a_3a_4x} &= -\sum_M C_{2M}^{a_1a_3}C_{2M}^{*\, a_2a_4}+\sum_M C_{1M}^{a_1a_3}C_{1M}^{*\, a_2a_4}+ 2C_{00}^{a_1a_3}C_{00}^{*\, a_2a_4}\, , \label{eq:f1t}\\
f^{a_1a_3x}f^{a_2a_4x} &= 2\sum_M C_{1M}^{a_1a_3}C_{1M}^{*\, a_2a_4}\,  . \label{eq:f2t}
\end{align}
These expressions can be substituted into Eqs.(\ref {gss1}), (\ref {gsu1}) and (\ref {gst1}), respectively, to obtain fully factorized forms of conformal block decompositions.

The block coefficients in the $(12\rightleftharpoons 34)_{{\mathfrak{2}}}$ channel, written in Eq.(\ref {gss1}), can be expressed as
\begin{align}
f^{a_1a_2x}&f^{a_3a_4x}a_m+f^{a_1a_3x}f^{a_2a_4x}\tilde{a}_m \nonumber\\
=&\sum_M C_{1M}^{a_1a_2}C_{1M}^{*a_3a_4}\Big[ \frac{m!\, \Gamma(-i\lambda_3+m)\Gamma(-i\lambda_4)}{\Gamma(i\lambda_2+2m-1)}+(-1)^m\frac{m!\,
\Gamma(-i\lambda_4+m)\Gamma(-i\lambda_3)}{\Gamma(i\lambda_2+2m-1)}\Big]\nonumber\\
&+(2C_{00}^{a_1a_2}C_{00}^{*a_3a_4}-\sum_M C_{2M}^{a_1a_2}C_{2M}^{*a_3a_4} )\nonumber\\
&~~\times\Big[ -\frac{m!\,\Gamma(-i\lambda_3+m)\Gamma(-i\lambda_4)}{\Gamma(i\lambda_2+2m-1)}+
(-1)^m\frac{m!\,\Gamma(-i\lambda_4+m)\Gamma(-i\lambda_3)}{\Gamma(i\lambda_2+2m-1)}\Big] \ , \end{align}\begin{align}
&f^{a_1a_2x}f^{a_3a_4x}b_m+f^{a_1a_3x}f^{a_2a_4x}\tilde{b}_m \nonumber\\
=&  \sum_M C_{1M}^{a_1a_2}C_{1M}^{*a_3a_4}\Big[ \frac{\Gamma(i\lambda_3)\Gamma(1-i\lambda_2+m)\Gamma(m+i\lambda_4)}{i\lambda_4\Gamma(-i\lambda_2)\Gamma(2m-i\lambda_2-1)}-(-1)^m\frac{\Gamma(i\lambda_4)\Gamma(1-i\lambda_2+m)\Gamma(m+i\lambda_3)}{i\lambda_3\Gamma(-i\lambda_2)\Gamma(2m-i\lambda_2-1)} \Big] \nonumber\\
&+(2C_{00}^{a_1a_2}C_{00}^{*a_3a_4}-\sum_M C_{2M}^{a_1a_2}C_{2M}^{*a_3a_4} ) \ , \nonumber\\
&~~\times\Big[ -\frac{\Gamma(i\lambda_3)\Gamma(1-i\lambda_2+m)\Gamma(m+i\lambda_4)}{i\lambda_4\Gamma(-i\lambda_2)\Gamma(2m-i\lambda_2-1)}-(-1)^m\frac{\Gamma(i\lambda_4)\Gamma(1-i\lambda_2+m)\Gamma(m+i\lambda_3)}{i\lambda_3\Gamma(-i\lambda_2)\Gamma(2m-i\lambda_2-1)} \Big]  \, . \label{eq:scoeffim}
\end{align}
Note that the coefficient of blocks which are even under $x\to -x$ in the small $x$ expansion are even (symmetric) under $3\leftrightarrow 4$, while odd blocks have odd (antisymmetric) coefficients. This property is necessary for $3\rightleftharpoons 4$
crossing symmetry to hold.

The block coefficients in the $(14\rightleftharpoons 32)_{{\mathfrak{2}}}$ channel, written in Eq.(\ref {gsu1}), can be expressed as
\begin{align}
f^{a_1a_2x}&f^{a_3a_4x}c_m+f^{a_1a_3x}f^{a_2a_4x}\tilde{c}_m \nonumber\\
=&\sum_M C_{1M}^{a_1a_4}C_{1M}^{*a_3a_2}\Big[ -\frac{m!\, \Gamma(-i\lambda_4-1)\Gamma(2+i\lambda_4)\Gamma(-i\lambda_3+m)}{\Gamma(-1+i\lambda_2)
\Gamma(1+i\lambda_4+2m)} \nonumber\\
&\qquad-(-1)^m \frac{m!\, \Gamma(-i\lambda_3)\Gamma(-i\lambda_4-1)\Gamma(2+i\lambda_4)
\Gamma(2-i\lambda_2+m)}{\Gamma(-1+i\lambda_2)\Gamma(2-i\lambda_2)\Gamma(1+i\lambda_4+2m)}
\Big]\nonumber\\
&+(2C_{00}^{a_1a_4}C_{00}^{*a_3a_2}-\sum_M C_{2M}^{a_1a_4}C_{2M}^{*a_3a_2})\Big[\frac{m!\, \Gamma(-i\lambda_4-1)\Gamma(2+i\lambda_4)\Gamma(-i\lambda_3+m)}{\Gamma(-1+i\lambda_2)
\Gamma(1+i\lambda_4+2m)} \nonumber\\
&\qquad-(-1)^m \frac{m!\, \Gamma(-i\lambda_3)\Gamma(-i\lambda_4-1)\Gamma(2+i\lambda_4)\Gamma(2-i\lambda_2+m)}{\Gamma(-1+i\lambda_2)
\Gamma(2-i\lambda_2)\Gamma(1+i\lambda_4+2m)}\Big]\ ,\end{align}\begin{align}
f^{a_1a_2x}&f^{a_3a_4x}d_m+f^{a_1a_3x}f^{a_2a_4x}\tilde{d}_m \nonumber\\
=& \sum_M C_{1M}^{a_1a_4}C_{1M}^{*a_3a_2}B(i\lambda_3,i\lambda_4-1)\nonumber\\
&\times\Big[ -\frac{\Gamma(-i\lambda_4+m-1)\Gamma(i\lambda_2+m-2)}{\Gamma(i\lambda_2-1)\Gamma(-i\lambda_4+2m-3)}+(-1)^m\frac{\Gamma(-i\lambda_4+m-1)\Gamma(i\lambda_3+m)}{\Gamma(1+i\lambda_3)\Gamma(-i\lambda_4+2m-3)}\Big]\nonumber\\
+&(2C_{00}^{a_1a_4}C_{00}^{*a_3a_2}-\sum_M C_{2M}^{a_1a_4}C_{2M}^{*a_3a_2})B(i\lambda_3,i\lambda_4-1)\nonumber\\
&\times\Big[ \frac{\Gamma(-i\lambda_4+m-1)\Gamma(i\lambda_2+m-2)}{\Gamma(i\lambda_2-1)\Gamma(-i\lambda_4+2m-3)}+(-1)^m\frac{\Gamma(-i\lambda_4+m-1)\Gamma(i\lambda_3+m)}{\Gamma(1+i\lambda_3)\Gamma(-i\lambda_4+2m-3)}\Big] \, . \label{eq:ucoeffim}
\end{align}
They are {\em not\/} related in any obvious way to the coefficients of
$(12\rightleftharpoons 34)_{{\mathfrak{2}}}$ blocks. The crossing symmetry
$2\rightleftharpoons 4$  interchanges incoming ($-\epsilon$) and outgoing ($+\epsilon$) operators, therefore it is not surprising that it is not manifest at the level of conformal blocks.

The block coefficients in the $(13\rightleftharpoons 42)_{{\mathfrak{2}}}$ channel, written in Eq.(\ref {gst1}), can be expressed as
\begin{align}
&f^{a_1a_2x}f^{a_3a_4x}e_m+f^{a_1a_3x}f^{a_2a_4x}\tilde{e}_m \nonumber\\
=& \sum_M C_{1M}^{a_1a_3}C_{1M}^{*a_4a_2}(-1)^m\Big[\frac{m!\, \Gamma(-i\lambda_3-1)\Gamma(2+i\lambda_3)\Gamma(-i\lambda_4+m)}{\Gamma(-1+i\lambda_2)
\Gamma(1+i\lambda_3+2m)} \nonumber\\
&\qquad + (-1)^m\frac{m!\,\Gamma(-i\lambda_4)
\Gamma(-i\lambda_3-1)\Gamma(2+i\lambda_3)\Gamma(2-i\lambda_2+m)}{\Gamma(-1+i\lambda_2)
\Gamma(2-i\lambda_2)\Gamma(1+i\lambda_3+2m)} \Big]\nonumber\\
&~~+(2C_{00}^{a_1a_3}C_{00}^{*a_4a_2}-\sum_M C_{2M}^{a_1a_3}C_{2M}^{*a_4a_2}) (-1)^m \Big[- \frac{m!\, \Gamma(-i\lambda_3-1)\Gamma(2+i\lambda_3)\Gamma(-i\lambda_4+m)}{\Gamma(-1+i\lambda_2)
\Gamma(1+i\lambda_3+2m)} \Big]\ ,\nonumber\\
&+(-1)^m\frac{m!\,\Gamma(-i\lambda_4)\Gamma(-i\lambda_3-1)\Gamma(2+i\lambda_3)
\Gamma(2-i\lambda_2+m)}{\Gamma(-1+i\lambda_2)\Gamma(2-i\lambda_2)
\Gamma(1+i\lambda_3+2m)}\ ,
\end{align}\begin{align}
f^{a_1a_2x}&f^{a_3a_4x}f_m+f^{a_1a_3x}f^{a_2a_4x}\tilde{f}_m \nonumber\\
=& \sum_M C_{1M}^{a_1a_3}C_{1M}^{*a_4a_2}B(i\lambda_4,i\lambda_3-1)(-1)^m\nonumber\\
&\times\Big[-\frac{\Gamma(-i\lambda_3+m-1)
\Gamma(i\lambda_2+m-2)}{\Gamma(i\lambda_2-1)\Gamma(-i\lambda_3+2m-3)} +(-1)^m\frac{\Gamma(-i\lambda_3+m-1)\Gamma(i\lambda_4+m)}{\Gamma(1+i\lambda_4)
\Gamma(-i\lambda_3+2m-3)}\Big]\nonumber\\
+&(2C_{00}^{a_1a_3}C_{00}^{*a_4a_2}-\sum_M C_{2M}^{a_1a_3}C_{2M}^{*a_4a_2})B(i\lambda_4,i\lambda_3-1)(-1)^m\nonumber\\
&\times\Big[\frac{\Gamma(-i\lambda_3+m-1)
\Gamma(i\lambda_2+m-2)}{\Gamma(i\lambda_2-1)\Gamma(-i\lambda_3+2m-3)}+(-1)^m \frac{\Gamma(-i\lambda_3+m-1)\Gamma(i\lambda_4+m)}{\Gamma(1+i\lambda_4)
\Gamma(-i\lambda_3+2m-3)}\Big]\, . \label{eq:tcoeffim}
\end{align}
As mentioned before, upon interchanging 3 and 4, the blocks of this channel match the blocks of
$(14\rightleftharpoons 32)_{{\mathfrak{2}}}$.
The coefficients have the same property:  they can be obtained
from Eq.(\ref{eq:ucoeffim}) by $3\leftrightarrow 4$ [modulo $(-1)^m$]. Here again,
these properties are necessary for $3\rightleftharpoons 4$ crossing symmetry -- while $2\rightleftharpoons 4$ is not manifest.

\section{Crossing symmetry and OPE}
In the previous section, we used analytic continuation to extend the single-valued correlator (\ref{gcorf}) to the entire complex plane. The conformal block decomposition was based on the implicit {\em assumption} of crossing symmetry between the channels. Namely, by construction,
\be
G_{32}^{41}(1-x,1-\bx)_{SV}=G_{34}^{21}(x,\bx)_{SV}\ ,\qquad G_{31}^{24}\Big(\frac{1}{x},\frac{1}{\bx}\Big)_{SV}=x^{2h_3}\bar{x}^{2\bh_3}
G_{34}^{21}(x,\bx)_{SV}\ .
\ee
Analyticity and crossing symmetry are two pillars of CFT, therefore they should be subjected to as many independent consistency checks as possible. A standard check is to prove the equivalence of OPEs extracted from distinct channels. Such self-consistency requirements are usually called the bootstrap conditions and often used for constructing CFT correlators. We can also compare OPEs with already known leading terms derived in Refs.\cite{Fan1903,Strominger1910}.\footnote{Note that in Ref.\cite{Fan1903}, gluon operators are normalized in a  different way than in Ref.\cite{Strominger1910}. Here, it is more convenient to use the normalization of Ref.\cite{Strominger1910}.}
The channel $(12\rightleftharpoons 34)_{{\mathfrak{2}}}$ probes directly (in the limit of $x\to 0$) the product
 of two outgoing gluon operators, $\phi^{+\epsilon}\phi^{+\epsilon}$,
  while   the channels $(14\rightleftharpoons 32)_{{\mathfrak{2}}}$ and
$(13\rightleftharpoons 42)_{{\mathfrak{2}}}$, in the limits of $x\to 1$ and  $x\to\infty$, respectively, probe, among other things, the OPEs
  $\phi^{+\epsilon}\phi^{-\epsilon}$ of one incoming gluon and one outgoing gluon.
The fact that, as pointed out in the previous section, $3\rightleftharpoons 4$ crossing symetry is manifest in conformal block decomposition
and that the blocks include gluon primary fields with $\Delta=1+i\lambda$, $J=\pm 1$
indicate that we are starting off on the right foot.
\subsection{OPE at the leading order}

We begin by extracting leading OPEs from the $(12\rightleftharpoons 34)_{{\mathfrak{2}}}$ channel, that is from the conformal block decomposition given in Eq.(\ref{gss1}). The block with $J=1$ and dimension $\Delta=1+i\lambda_3+i\lambda_4$ originates from the $m=1$ term  in the second line.
As  $x=(z_{12}z_{34})(z_{13}z_{24})^{-1}\to 0$, it is the leading term with $K_{34}^{21}\big[1+\frac{i\lambda_3}{2} +\frac{i\lambda_4}{2} ,\frac{i\lambda_3}{2}+\frac{i\lambda_4}{2} \big](x,\bx)\sim 1/x$. According to Eq.(\ref{eq:scoeffim}), the coefficient is
\begin{align}
&f^{a_1a_2x}f^{a_3a_4x} b_1\, +\,f^{a_1a_3x}f^{a_2a_4x}\tilde{b}_1 = f^{a_1a_2x}f^{a_3a_4x}(1-i\lambda_2)B(i\lambda_3,i\lambda_4) \nonumber\\[1mm]
&\quad=\frac{-f^{a_1a_2x}}{\Gamma(2-\Delta_2-\Delta_3-\Delta_4)}B(\Delta_2-1, 2-\Delta_2-\Delta_3-\Delta_4)\big[f^{a_3a_4x}B(\Delta_3-1,\Delta_4-1)\big] ,
 \label{eq:pspin-1}
\end{align}
where $\Delta_i=1+i\lambda_i$. The constant $[\Gamma(2-\Delta_2-\Delta_3-\Delta_4)]^{-1}=[\Gamma(\Delta_1-2)]^{-1}$ is to be understood as $\lim_{\lambda_1\to 0}[\Gamma(-1+i\lambda_1)]^{-1}$ and is formally zero in this limit.
It is present in all blocks and can be incorporated into the definition of the (soft) shadow field.

The block coefficient (\ref{eq:pspin-1}) can be factorized into the OPE coefficients in the following way. In the limit of $x\to 0$ ($z_3\to z_4$), the leading OPE term is \cite{Fan1903,Strominger1910}
\begin{align}
\phi_{\Delta_3,+}^{a_3,+\epsilon}(z_3,\bz_3)\phi_{\Delta_4,+}^{a_4,+\epsilon}(z_4,\bz_4) &\sim \frac{-i f^{a_3a_4x}}{z_{34}} B(\Delta_3-1,\Delta_4-1)\phi_{\Delta_3+\Delta_4-1,+}^{x,+\epsilon}(z_4,\bz_4)\, .\label{eq:OoutOout1}
\end{align}
The coefficient of the above OPE accounts for the last factor on the r.h.s.\ of
Eq.(\ref{eq:pspin-1}), enclosed in square brackets. In the next step, we use the OPE of the resultant operator with the remaining gluon, incoming at $z_2$ \cite{Strominger1910}:
\begin{align}
\phi_{\Delta_2,-}^{a_2,-\epsilon}(z_2,&\bz_2) \phi_{\Delta_3+\Delta_4-1,+}^{x,+\epsilon}(z_4,\bz_4)\nonumber\\[1mm]
& \sim ~\frac{i f^{a_2xy}}{\bz_{24}}  \Big[B(\Delta_2-1, 2-\Delta_2-\Delta_3-\Delta_4)\phi^{y,+\epsilon}_{\Delta_2+\Delta_3+\Delta_4-2,+}(z_4,\bz_4)
\nonumber\\[1mm]
& ~~~~~-B(\Delta_3+\Delta_4,2-\Delta_2-\Delta_3-\Delta_4)\phi^{y,-\epsilon}_{\Delta_2
+\Delta_3+\Delta_4-2,+}(z_4,\bz_4)\Big]+\dots, \label{eq:OinOout1}
\end{align}
where  we omitted negative helicity gluon operators because their two-point functions with the (positive helicity) shadow field vanish due to conformal invariance. At this point, we are left with the correlator of the shadow field with one gluon operator. The coordinate dependence of this correlator is fixed by conformal invariance, {\em c.f}.\ Eq.(I.2.8). Here, in order to produce the remaining factor in the block coefficient (\ref{eq:pspin-1}), we need to make one additional assumption, that non-vanishing correlators must involve
 one incoming and one outgoing operator:
\begin{align}
\langle\tilde\phi_{\tilde\D_1=1,+}^{a_1,-\epsilon}(z_{1},\zbar_{1})\,
\phi^{y,-\epsilon}_{\Delta_2+\Delta_3+\Delta_4-2,+}(z_4,\bz_4)\rangle &=0\ ,\label{zero2pt}\\[1mm]
\langle\tilde\phi_{\tilde\D_1=1,+}^{a_1,-\epsilon}(z_{1},\zbar_{1})\,
\phi^{y,+\epsilon}_{\Delta_2+\Delta_3+\Delta_4-2,+}(z_4,\bz_4)\rangle &=
\langle\tilde\phi_{\tilde\D_1=1,+}^{a_1,-\epsilon}(z_{1},\zbar_{1})\,
\phi^{y,+\epsilon}_{2-\Delta_1,+}(z_4,\bz_4)\rangle\nonumber\\[1mm] &
=\frac{-\delta^{a_1y}}{\Gamma(\Delta_1-2)z_{14}^2} \label{eq:2pt}\ .
\end{align}
Under this assumption, from the r.h.s.\ of Eq.(\ref{eq:OinOout1}), only the first term contributes,
and the product of the OPE coefficients yields the block coefficient (\ref{eq:pspin-1}). Indeed, after substituting the sequence (\ref{eq:OoutOout1})-(\ref{eq:2pt}) into Eq.(\ref{gcorf}), we obtain the leading $1/x$ term with the right coefficient (\ref{eq:pspin-1}).

We now proceed to the $(14\rightleftharpoons 32)_{{\mathfrak{2}}}$ channel, with the conformal block decomposition given in Eq.(\ref{gsu1}). Here, we encounter a gluon block with $\Delta=1+i\lambda_2+i\lambda_3$ and $J=-1$, which originates from $m=1$ in the second line of Eq.(\ref{gsu1}). As  $1-x=(z_{14}z_{23})(z_{13}z_{24})^{-1}\to 0$, it is the leading term with $K_{32}^{41}\big[\frac{i\lambda_2}{2} +\frac{i\lambda_3}{2} ,1+\frac{i\lambda_2}{2}+\frac{i\lambda_3}{2} \big](1-x,1-\bx)\sim 1/(1-x)$. According to Eq.(\ref{eq:ucoeffim}), the coefficient is
\begin{align}
&f^{a_1a_2x}f^{a_3a_4x} d_1\, +\,f^{a_1a_3x}f^{a_2a_4x}\tilde{d}_1 =
f^{a_4a_1x}f^{a_2a_3x}(1+i\lambda_4)B(i\lambda_3,-i\lambda_2-i\lambda_3-1)
 \nonumber\\[1mm]
&\quad=\frac{f^{a_1a_4x}}{\Gamma(2-\Delta_2-\Delta_3-\Delta_4)}B(\Delta_2+\Delta_3-2, 2-\Delta_2-\Delta_3-\Delta_4)\nonumber\\ &\qquad\qquad\qquad\qquad\qquad\qquad\times\big[f^{a_2a_3x}B(\Delta_3-1,1-\Delta_2-\Delta_3)\big] .
 \label{eq:pspin+1u}
\end{align}

The block coefficient (\ref{eq:pspin+1u}) can be factorized into OPE coefficients in the following way. In the limit of $x\to 1$ ($z_2\to z_3$), the leading OPE terms are \cite{Strominger1910}
\begin{align}
\phi_{\Delta_2,-}^{a_2,-\epsilon}(z_2,\bz_2)\phi_{\Delta_3,+}^{a_3,+\epsilon}(z_3,\bz_3)
& \sim \frac{-i f^{a_2a_3x}}{z_{23}}  \Big[B(\Delta_3-1, 1-\Delta_2-\Delta_3)\phi^{x,-\epsilon}_{\Delta_2+\Delta_3-1,-}(z_3,\bz_3)\nonumber\\
&\qquad-B(\Delta_2+1,1-\Delta_2-\Delta_3)\phi^{x,+\epsilon}_{\Delta_2+\Delta_3-1,-}(z_3,\bar{z}_3)\Big]
\nonumber\\
& + \frac{-i f^{a_2a_3x}}{\bar{z}_{23}} \Big[B(\Delta_3+1, 1-\Delta_2-\Delta_3)\phi^{x,-\epsilon}_{\Delta_2+\Delta_3-1,+}(z_3,\bz_3)\nonumber\\
&\qquad-B(\Delta_2-1,1-\Delta_2-\Delta_3)\phi^{x,+\epsilon}_{\Delta_2+\Delta_3-1,+}(z_3,\bar{z}_3)\Big]
\, , \label{eq:OinOout2}\end{align}
Note that this OPE contains both $1/z_{23}\sim 1/(1-x)$ and $1/\bz_{23}\sim 1/(1-\bx)$ poles, while the latter ones are absent in the conformal blocks of Eq.(\ref{gsu1}). Hence, we need to show two things:
that the conformal block $\sim 1/(1-x)$ emerging from the $1/z_{23}$ terms in Eq.(\ref{eq:OinOout2}) has the right coefficient and that the $1/\bz_{23}$ terms do not contribute to the four-point correlator.

We begin with the $1/z_{23}$ terms of Eq.(\ref{eq:OinOout2}). The OPEs of the respective operators with the gluon outgoing at $z_4$ are \cite{Fan1903,Strominger1910}:
\begin{align}\phi_{\Delta_4,+}^{a_4,+\epsilon}(z_4,\bz_4) &
\phi_{\Delta_2+\Delta_3-1,-}^{x,-\epsilon}(z_2,\bz_2) \nonumber\\
& \sim \frac{-i f^{a_4xy}}{\bz_{24}}  \Big[B(\Delta_2+\Delta_3-2, 2-\Delta_2-\Delta_3-\Delta_4)\phi^{y,+\epsilon}_{\Delta_2+\Delta_3+\Delta_4-2,+}(z_4,\bz_4)\nonumber\\
&-B(\Delta_4+1, 2-\Delta_2-\Delta_3-\Delta_4)\phi^{y,-\epsilon}_{\Delta_2+\Delta_3+\Delta_4-2,+}(z_4,\bz_4)\Big]+\dots , \label{eq:OinOout3}\\
 \phi_{\Delta_4,+}^{a_4,+\epsilon}(z_4,\bz_4)&
\phi_{\Delta_2+\Delta_3-1,-}^{x,+\epsilon}(z_2,\bz_2)\nonumber \\ &\sim \frac{i f^{a_4xy}}{\bz_{24}} B(\Delta_2+\Delta_3-2,\Delta_4+1)\phi^{y,+\epsilon}_{\Delta_2+\Delta_3+\Delta_4-2,+}(z_4,\bz_4) \, .\label{eq:OoutOout-+}
\end{align}
Here again, we omitted negative helicity operators on the r.h.s.\ because their two-point functions with the shadow field are vanishing. Furthermore, the two-point correlator (\ref{eq:2pt}) contains the normalization constant $[\Gamma(\Delta_1-2)]^{-1}$, which is formally zero in the limit of $\lambda_1=0$. Therefore, the second OPE (\ref{eq:OoutOout-+}), which is finite in this limit, will not contribute to the four-point function. Taking into account that the two-point correlator with an incoming field is zero, we conclude that the leading $1/(1-x)$ pole originates from the first terms of Eqs.(\ref{eq:OinOout2}) and (\ref{eq:OinOout3}). In this way, we obtain the coefficient (\ref{eq:pspin+1u}).

It remains to be proven that the $1/\bz_{23}$ pole terms of the OPE (\ref{eq:OinOout2}) do not contribute to the four-point function. The OPEs of the respective operators with the gluon outgoing at $z_4$ are \cite{Fan1903,Strominger1910}:
\begin{align}\phi_{\Delta_4,+}^{a_4,+\epsilon}(z_4,\bz_4) &
\phi_{\Delta_2+\Delta_3-1,+}^{x,-\epsilon}(z_2,\bz_2) \nonumber\\
& \sim \frac{-i f^{a_4xy}}{z_{24}}  \Big[B(\Delta_2+\Delta_3-2, 4-\Delta_2-\Delta_3-\Delta_4)\phi^{y,+\epsilon}_{\Delta_2+\Delta_3+\Delta_4-2,+}(z_4,\bz_4)\nonumber\\
&-B(\Delta_4-1, 4-\Delta_2-\Delta_3-\Delta_4)\phi^{y,-\epsilon}_{\Delta_2+\Delta_3+\Delta_4-2,+}(z_4,\bz_4)\Big]\, , \label{eq:OinOout4}\\
 \phi_{\Delta_4,+}^{a_4,+\epsilon}(z_4,\bz_4)&
\phi_{\Delta_2+\Delta_3-1,+}^{x,+\epsilon}(z_2,\bz_2)\nonumber\\ &\sim \frac{i f^{a_4xy}}{z_{24}} B(\Delta_2+\Delta_3-2,\Delta_4-1)
\phi^{y,+\epsilon}_{\Delta_2+\Delta_3+\Delta_4-2,+}(z_4,\bz_4) \, .\label{eq:OoutOout++}
\end{align}
All these OPEs are finite in the limit of $\lambda_1=0$, therefore they do not contribute to the four-point function, for the same reason as (\ref{eq:OoutOout-+}).

{}Finally, we turn to the $(13\rightleftharpoons 42)_{{\mathfrak{2}}}$ channel, with the conformal block decomposition given in Eq.(\ref{gst1}). Here, we encounter a gluon block with $\Delta=1+i\lambda_2+i\lambda_4$ and $J=-1$, which originates from $m=1$ in the second line of Eq.(\ref{gst1}). As  $1/x=(z_{13}z_{24})(z_{12}z_{34})^{-1}\to 0$, it is the leading term with $K_{31}^{24}\big[\frac{i\lambda_2}{2} +\frac{i\lambda_4}{2} ,1+\frac{i\lambda_2}{2}+\frac{i\lambda_4}{2} \big]\big(\frac{1}{x},\frac{1}{\bx}\big)\sim x$. According to Eq.(\ref{eq:tcoeffim}), the coefficient is
\begin{align}
&f^{a_1a_2x}f^{a_3a_4x} f_1\, +\,f^{a_1a_3x}f^{a_2a_4x}\tilde{f}_1 =
f^{a_1a_3x}f^{a_2a_4x}(1+i\lambda_3)B(i\lambda_4,-i\lambda_2-i\lambda_4-1)
 \nonumber\\[1mm]
&\qquad=\frac{-f^{a_1a_3x}}{\Gamma(2-\Delta_2-\Delta_3-\Delta_4)}B(\Delta_2+\Delta_4-2, 2-\Delta_2-\Delta_3-\Delta_4)\nonumber\\ &\quad\qquad\qquad\qquad\qquad\qquad\qquad\times
\big[f^{a_2a_4x}B(\Delta_4-1,1-\Delta_2-\Delta_4)\big] .
 \label{eq:pspin+1t}
\end{align}
We see that it is equal to  (\ref{eq:pspin+1u}) with $3\leftrightarrow 4$, therefore in order to factorize it into OPE coefficients,  we can repeat the same sequence of steps as in the $(14\rightleftharpoons 32)_{{\mathfrak{2}}}$ channel. We conclude that crossing symmetry is consistent with  OPE at the leading order. The leading OPE terms, derived in Refs.\cite{Fan1903} and \cite{Strominger1910}, yield the same coefficients of leading blocks as the single-valued correlator. Therefore, the single-valued completion restores agreement with OPE and crossing symmetry at the same time.

\subsection{Examples of non-leading OPE terms}

In the first example, we consider $\Delta=2+i\lambda_3+i\lambda_4$, $J=2$ primaries that appear in the $(12\rightleftharpoons 34)_{{\mathfrak{2}}}$ channel. The corresponding block
$K_{34}^{21}\big[2+\frac{i\lambda_3}{2} +\frac{i\lambda_4}{2} ,\frac{i\lambda_3}{2}+\frac{i\lambda_4}{2} \big]\sim 1$ when $x\to 0$. According to Eq.(\ref{eq:scoeffim}), the coefficient is
\begin{align}
f^{a_1a_2x}&f^{a_3a_4x}b_2+f^{a_1a_3x}f^{a_2a_4x}\tilde{b}_2 \nonumber\\
=&\sum_{M}C_{1M}^{a_1a_2}C_{1M}^{*\, a_3a_4} \,i\lambda_2\big[ B(i\lambda_4,1+i\lambda_3)-B(i\lambda_3,1+i\lambda_4)\big] \nonumber\\
&+(-\sum_{M}C_{2M}^{a_1a_2}C_{2M}^{*\, a_3a_4}+2C_{00}^{a_1a_2}C_{00}^{*a_3a_4})(i\lambda_2-2)B(i\lambda_3,i\lambda_4) \, .\label{coe2}
\end{align}
Note that the above expression is symmetric under $3\leftrightarrow 4$ because $I=1$ Clebsch-Gordan coefficients are antisymmetric under this transposition, while $I=0$ and $I=2$ are symmetric. The conformal block coefficient of $I=0$ group singlet,
\begin{align} 2&C_{00}^{a_1a_2}C_{00}^{*a_3a_4}(i\lambda_2-2)B(i\lambda_3,i\lambda_4) \nonumber\\
&~=-\widetilde{C} \sqrt 2C_{00}^{a_1a_2}B(\Delta_2-2,2-\Delta_2-\Delta_3-\Delta_4)\big[\sqrt 2C_{00}^{*a_3a_4}B(\Delta_3-1,\Delta_4-1)\big]
\, ,
\end{align}
where $\widetilde{C}\equiv -[\Gamma(\Delta_1-2)]^{-1}$,
can be obtained from the following OPE terms
\begin{align}
\phi_{\Delta_3,+}^{a_3,+\epsilon}(z_3,\bz_3) & \phi^{a_4,+\epsilon}_{\Delta_4,+}(z_4,\bz_4) \sim -\sqrt{2}C_{00}^{*a_3a_4} B(\Delta_3-1,\Delta_4-1)
O^{00,+\epsilon}_{\Delta_3+\Delta_4,+2}(z_4,\bz_4)+\dots, ~~~~ \label{eq:pspin2singlet1}\\[1mm]
\phi_{\Delta_2,-}^{a_2,-\epsilon}(z_2,\bz_2)&O^{00,+\epsilon}_{\Delta_3+\Delta_4,+2}(z_4,\bz_4) \nonumber\\
&\sim  \frac{\sqrt{2}C_{00}^{a_2x}}{z_{24}\bz_{24}}B(\Delta_2-2,2-\Delta_2-\Delta_3-\Delta_4)\phi^{x, +\epsilon}_{\Delta_2+\Delta_3+\Delta_4-2,-}(z_4,\bz_4)+\dots,\label{eq:pspin2singlet2}
\end{align}
where $O^{00,+\epsilon}_{\Delta,+2}$ is an outgoing primary operator with $I=0, J=2$.
The coefficient of $J=2$, $I=2$ quintuplet follows from similar OPEs. On the other hand, the coefficient of $J=2$, $I=1$ triplet,
\begin{align} \sum_{M}&C_{1M}^{a_1a_2}C_{1M}^{*\, a_3a_4} \,i\lambda_2\big[ B(i\lambda_4,1+i\lambda_3)-B(i\lambda_3,1+i\lambda_4)\big]\nonumber\\ &=
-\widetilde{C}\sum_{M}C_{1M}^{a_1a_2}B(\Delta_2, 2-\Delta_2-\Delta_3-\Delta_4)C^{*a_3a_4}_{1M}
\big[B(\Delta_4-1,\Delta_3)-B(\Delta_3-1,\Delta_4)\big],
\end{align}
can be obtained from
\begin{align}
\phi_{\Delta_3,+}^{a_3,+\epsilon}&(z_3,\bz_3) \phi^{a_4,+\epsilon}_{\Delta_4,+}(z_4,\bz_4)\nonumber\\ &\sim -\sum_M C^{*a_3a_4}_{1M}\big[B(\Delta_4-1,\Delta_3)-B(\Delta_3-1,\Delta_4)\big]O^{1M,+
\epsilon}_{\Delta_3+\Delta_4,+2}(z_4,\bz_4)+\dots,  \label{eq:pspin2tri1}\\
\phi_{\Delta_2,-}^{a_2,-\epsilon}&(z_2,\bz_2)O^{1M,+\epsilon}_{\Delta_3+\Delta_4,+2}(z_4,\bz_4) \nonumber\\
&\sim \frac{C_{1M}^{a_2x}}{z_{24}\bz_{24}}B(\Delta_2, 2-\Delta_2-\Delta_3-\Delta_4)\phi^{x,+\epsilon}_{\Delta_2+\Delta_3+\Delta_4-2,+}(z_4,\bz_4)
+\dots\label{eq:pspin2tri2}
\end{align}
Note that $3\leftrightarrow 4$ symmetry of the conformal block coefficient (\ref{coe2}) is necessary for self-consistent OPE of $ \phi_{\Delta_3,+}^{a_3,+\epsilon} \phi^{a_4,+\epsilon}_{\Delta_4,+}$.

In the second example, we consider $\Delta=2+i\lambda_3+i\lambda_4$, $J=0$ primaries that appear in the $(14\rightleftharpoons 32)_{{\mathfrak{2}}}$ channel. The corresponding block
$K_{32}^{41}\big[1+\frac{i\lambda_2}{2} +\frac{i\lambda_3}{2} ,1+\frac{i\lambda_2}{2}+\frac{i\lambda_3}{2} \big]\sim 1$ when $x\to 1$.
According to Eq.(\ref{eq:ucoeffim}), its coefficient is
\begin{align}
f^{a_1a_2x}&f^{a_3a_4x} d_2+f^{a_1a_3x}f^{a_2a_2x} \tilde{d}_2 \nonumber\\
=&(\sum_M C_{2M}^{a_1a_4}C_{2M}^{*a_3a_2} -2 C_{00}^{a_1a_4}C_{00}^{*a_3a_2})B(i\lambda_3,i\lambda_4-1)i\lambda_4 \nonumber\\
&+\sum_M C_{1M}^{a_1a_4}C_{1M}^{*a_3a_2} B(i\lambda_3,i\lambda_4-1)(2+i\lambda_3-i\lambda_2) \, . \label{eq:d2dt2}
\end{align}
The conformal block coefficient of $I=0$ group singlet,
\begin{align} -2 & C_{00}^{a_1a_4}C_{00}^{*a_3a_2}B(i\lambda_3,i\lambda_4-1)
i\lambda_4\nonumber\\[1mm]&=
-\widetilde{C}\sqrt 2 C_{00}^{a_1a_4} B(\Delta_2+\Delta_3-1,2-\Delta_2-\Delta_3-\Delta_4)
\sqrt2 C_{00}^{*a_3a_2}B(\Delta_3-1,1-\Delta_2-\Delta_3)
\end{align}
can be factorized into OPE coefficients in a similar way as in the $(12\rightleftharpoons 34)_{{\mathfrak{2}}}$ channel:
\begin{align}
\phi_{\Delta_3,+}^{a_3,+\epsilon}(z_3,&\bz_3)\phi_{\Delta_2,-}^{a_2,-\epsilon}(z_2,\bz_2) \sim -\sqrt{2}C_{00}^{*a_3a_2}B(\Delta_3-1,1-\Delta_2-\Delta_3)
O^{00,-\epsilon}_{\Delta_2+\Delta_3,0}(z_2,\bz_2) +\dots,\label{eq:pspin0single1} \\
\phi_{\Delta_4,+}^{a_4,+\epsilon}(z_4,&\bz_4)O^{00,-\epsilon}_{\Delta_2+\Delta_3,0}(z_2,\bz_2) \nonumber\\[1mm]
&\sim \frac{\sqrt{2}C_{00}^{a_4x}}{z_{24}\bz_{24}}B(\Delta_2+\Delta_3-1,2-
\Delta_2-\Delta_3-\Delta_4)\phi^{x,+\epsilon}_{\Delta_2+\Delta_3+\Delta_4-2,+}(z_2,\bz_2)+\dots \label{eq:pspin0single2}
\end{align}
The case of $I=1$ triplet, however, is more complicated because its coefficient
\be \sum_M C_{1M}^{a_1a_4}C_{1M}^{*a_3a_2} B(i\lambda_3,i\lambda_4-1)(2+i\lambda_3-i\lambda_2)
\ee
contains the factor $(2+i\lambda_3-i\lambda_2)$, which cannot be written as a Beta function that could represent a single OPE coefficient, as it was in the case of all previous examples. It can be written though as a sum of (at least three) Beta functions, each of them attributed to an OPE coefficient of a distinct primary with the same dimensions, spin, and isospin. The conformal block spectrum must contain a certain degree of degeneracy in order to allow for the splitting of block coefficients into sums of products of OPE coefficients associated with distinct Verma modules, all with the same conformal dimension and spin and in the same group representation. This is an indication of another quantum number that could possibly distinguish between such modules.

As pointed out in Ref.\cite{Strominger1910}, the OPE coefficients of CCFT are constrained by supertranslational invariance under which  primary operators transform as
\be
\delta_{\cal{P}} O_{\Delta}^{\pm\epsilon} = \pm O^{\pm\epsilon}_{\Delta+1} \, , \label{eq:calPO}
\ee
where $\cal P$ is the supertranslation generator \cite{Stieberger1812}.
{}For an OPE of the form \be
O_{\Delta_1}^{\pm\epsilon}O_{\Delta_2}^{\pm\epsilon}\sim C(\Delta_1,\Delta_2)\,f(z_{12},\bz_{12}) \, O_{\Delta_1+\Delta_2+n}^{\pm\epsilon}\ ,\ee
this constraint reads
\be \pm C(\Delta_1+1,\Delta_2)\pm C(\Delta_1,\Delta_2+1)=\pm C(\Delta_1,\Delta_2)\ \ee
and  must be respected when extracting OPEs from conformal blocks. It is easy to check that all OPE coefficients written above satisfy this constraint.

\section{Integral representations of the single-valued correlator}
In the Coulomb gas formulation of minimal models, the correlators of primary fields (vertex operators) can be represented as complex integrals \cite{DiF,dots} over the positions of ``charge-screening'' vertices. The single-valued correlator (\ref{repf}) bears a striking resemblance to the four-point correlator in minimal models with Verma modules degenerating at level 2. Therefore, we expect it to be represented by a single integral. In this section, we construct this integral by following the classic approach of Dotsenko and Fateev \cite{Dotsenko:1984ad,Dotsenko:1984nm,dots}. The integrand involves insertions of charged vertices on a (celestial) sphere, therefore the correlator has a form similar to a Koba-Nielsen amplitude in closed string theory, with Coulomb charges (related to conformal dimensions) mapped to momenta of external string states and a celestial sphere mapped to a string world-sheet.\footnote{This mapping appears naturally in the large conformal dimension limit of Mellin-transformed string amplitudes
\cite{Stieberger:2018edy}.} It is well known that closed string amplitudes can be obtained
from open string amplitudes -- given by integrals iterated over the segments of $\mathbf{R}^{\!\mathbf{1}}$ --
by the so-called single-valued projection. By following this approach, we will rewrite the correlator (\ref{repf}) as a single-valued projection of a real-line integral.
\subsection{Complex integrals in Dotsenko-Fateev form}
Our goal is to express the correlator (\ref{repf}) in terms of single-valued complex integrals.
The integrals considered by Dotsenko and Fateev \cite{Dotsenko:1984ad,Dotsenko:1984nm,dots} in the context of minimal models have the form
\begin{equation}\label{complexI}
{\cal I}(x,\bar x)=\int d^2w\ w^{\hat a+a}\ \bar w^{\hat a+\bar a}\ (w-1)^{\hat b+b}\ (\bar w-1)^{\hat b+\bar b}\ (w-x)^{\hat c+c}\ (\bar w-\bar x)^{\hat c+\bar c}\ ,
\end{equation}
with the parameters $a,b,c,\bar a,\bar b,\bar c\in \IZ$, where $\IZ$ is the set of integers, and non-integer $\hat a,\hat b,\hat c\notin\IZ$.
The integral \req{complexI} can be evaluated through analytic continuation by disentangling holomorphic and antiholomorphic parts \cite{Dotsenko:1984ad,Dotsenko:1984nm,dots}:\footnote{This procedure is also known as the Kawai-Lewellen-Tye (KLT) method \cite{Kawai:1985xq}. Holomorphic and anti-holomorphic coordinates $w,\bar w$ become independent coordinates $\xi,\eta$, respectively. The latter are integrated from $-\infty$ to $+\infty$ subject to phase factors rendering the integrand single-valued when the branch points $w\equiv\xi=0,x,1$ and $\bar w\equiv\eta=0,\bar x,1$ are crossed.}
\begin{align}
{\cal I}(x,\bar x)&=\frac{s(\hat b)s(\hat a+\hat b+\hat c)}{s(\hat a+\hat c)}\
{\cal I}_1(\hat a+a,\hat b+b,\hat c+c;x)\ {\cal I}_1(\hat a+\bar a,\hat b+\bar b,\hat c+\bar c;\bar x)\nonumber\\
&+\ (-1)^{b+\bar b+c+\bar c}\ \frac{s(\hat a)s(\hat c)}{s(\hat a+\hat c)}\  {\cal I}_2(\hat a+a,\hat b+b,\hat c+c;x)\ {\cal I}_2(\hat a+\bar a,\hat b+\bar b,\hat c+\bar c;\bar x)\ ,\label{i12}
\end{align}
with $s(x)\equiv \sin(\pi x)$ and
\begin{align}
 {\cal I}_1(a,b,c;x)&=\int_1^\infty dw\ w^a\ (w-1)^b\ (w-x)^c\nonumber\\
 &=B(-a-b-c-1,b+1)\ {}_2F_1\left({-c,-a-b-c-1\atop -a-c};x\right)\ ,\label{Int1}\\
{\cal I}_2(a,b,c;x)&=\int_0^x dw\ w^a\ (1-w)^b\ (x-w)^c\nonumber\\
&=x^{1+a+c}\ B(a+1,c+1)\ {}_2F_1\left({-b,1+a\atop a+c+2};x\right)\ .\label{Int2}
\end{align}
To get explicit relations, we restricted ourselves to
$0<x<1$. The general case is obtained by analytic continuation.

In order to recast the correlator (\ref{repf}) in a form similar to (\ref{i12}), we observe that the single-valued combination (\ref{svcom}) derived in Section 2 can be written as
\begin{align}\label{castG}S_1(x)&\bar{I}_1(\bar{x}) +S_2(x)\bar{I}_2(\bar{x})\nonumber\\
&=\frac{1}{x(1-x)} \left\{  (1-i\lambda_2) B(i\lambda_3,i\lambda_4) \, ~ _2F_1\left({-1, i\lambda_3\atop 1-i\lambda_2}; x\right) {}_2F_1\left({1,\, i\lambda_3\atop -i\lambda_2};\bar{x}\right)\right. \nonumber \\
& ~~~\left.+B(1-i\lambda_3,-i\lambda_4)x^{i\lambda_2} {}_2F_1\left({-1+i\lambda_2,-i\lambda_4\atop 1+i\lambda_2};x\right)\bar{x}^{1+i\lambda_2}{}_2F_1\left({2+i\lambda_2,1-i\lambda_4\atop 2+i\lambda_2};\bar x\right)\right\}.
\end{align}
Now it becomes clear that
\begin{equation}\label{Gcomplex}
S_1(x)\bar{I}_1(\bar{x}) +S_2(x)\bar{I}_2(\bar{x}) = \frac{1}{\pi}\ (1+i\lambda_2)\ B(i\lambda_3,i\lambda_4) \
\frac{{\cal I}(x,\bar x)}{x(1-x)}\ ,
\end{equation}
with the following parameters:
\begin{equation}\label{abc}
{\cal I}(x,\bar x):  \left\{\begin{array}{ccccccc}
\hat a=-i\lambda_4\ &,&\ a=-1\ &,&\ \bar a=0\ ,\\
\hat b=-i\lambda_2\ &,&\ b=1\ &,&\ \bar b=-2\ ,\\
\hat c=-i\lambda_3\ &,&\ c=0\ &,&\ \bar c=0\ .\\
\end{array}\right.
\end{equation}
Similarly,
\begin{align}\label{casttG}\tilde S_1(x)&\bar{I}_1(\bar{x}) +\tilde S_2(x)\bar{I}_2(\bar{x}) \nonumber\\ &= -\frac{1}{1-x}\left\{(2-i\lambda_2)B(i\lambda_3,i\lambda_4) \, ~_2F_1\lf({-1,\, 1+i\lambda_3\atop  2-i\lambda_2}; x\ri)
{}_2F_1\left({1, i\lambda_3\atop -i\lambda_2};\bar{x}\right) \right.\nonumber\\
&\lf.+B(-i\lambda_3,-i\lambda_4)\ x^{-1+i\lambda_2}\, _2F_1\left({-2+i\lambda_2, -i\lambda_4\atop i\lambda_2}; x\right)\bar{x}^{1+i\lambda_2}{}_2F_1\left({2+i\lambda_2,1-i\lambda_4\atop 2+i\lambda_2};\bar x\right)\right\}.
\end{align}
By comparing this expression with Eq.(\ref{i12}), we see that
\begin{equation}\tilde S_1(x)\bar{I}_1(\bar{x}) +\tilde S_2(x)\bar{I}_2(\bar{x})
= -\frac{1}{\pi}\ (1+i\lambda_2)\ B(i\lambda_3,i\lambda_4) \
\frac{ {\cal I}^{ \prime}(x,\bar x)}{1-x}\ ,\label{stcom}
\end{equation}
with the assignments:
\begin{equation}
{\cal I}^\prime(x,\bar x): \left\{\begin{array}{ccccccc}
\hat a=-i\lambda_4\ &,&\ a=-1\ &,&\ \bar a=0\ ,\\
\hat b=-i\lambda_2\ &,&\ b=2\ &,&\ \bar b=-2\ ,\\
\hat c=-i\lambda_3\ &,&\ c=-1\ &,&\ \bar c=0\ .\\
\end{array}\right.\label{assign1}
\end{equation}
In this way, we obtain the following integral representation:
\begin{align}
&G_{34}^{21}(x,\bar{x})_{SV}=\frac{1}{\pi}\ (1+i\lambda_2)\ B(i\lambda_3,i\lambda_4)\frac{1}{x(1-x)}\,\times \nonumber\\
&~~\left\{f^{a_1a_2b}f^{a_3a_4b}\!\int\!\! d^2w\, w^{-1-i\lambda_4} \bar w^{-i\lambda_4} (w-1)^{1-i\lambda_2} (\bar w-1)^{-2-i\lambda_2} (w-x)^{-i\lambda_3} (\bar w-\bar x)^{-i\lambda_3}\right.
\nonumber \\[1mm]
& ~-x\!\left.f^{a_1a_3b}f^{a_2a_4b}\!\!
\int\!\! d^2w\, w^{-1-i\lambda_4} \bar w^{-i\lambda_4}(w-1)^{2-i\lambda_2} (\bar w-1)^{-2-i\lambda_2} (w-x)^{-1-i\lambda_3}(\bar w-\bar x)^{-i\lambda_3}\right\} .\label{repint}\end{align}
This correlator has a form of a ``heterotic'' Coulomb gas correlator: the holomorphic factors have exponents  differing by integers from the antiholomorphic ones.
\subsection{Complex integrals as single-valued projections}
The complex integral \req{complexI} can be expressed as a linear combination (with rational coefficients  in $\hat a,\hat b, \hat c$)  of a basis of real (iterated) integrals subject to the single--valued projection $\sv$.\footnote{See Appendix B for a short review and references.} For a given choice of integers $a,b,c,\bar a,\bar b, \bar c$, this is achieved by first expressing  \req{complexI}  in terms of the four basis elements \req{Jmatrix} and then applying the relation \req{svRel}.

For the two cases \req{abc} and \req{assign1} (with  $\bar a,\bar c=0,\bar b=-2$), by partial integrating and partial fractioning in the anti-holomorphic part, we may first cast \req{complexI} into a linear combination of two complex integrals:
\be\label{startI}
\Ic(x,x)_{\bar a,\bar c=0\atop \bar b=-2}=\fc{\hat a\bar x}{1-\hat b}\ \Ic(x,x)_{\bar b=0\atop \bar a,\bar c=-1}+\fc{\hat a(1-\bar x)+\hat c}{1-\hat b}\ \Ic(x,x)_{\bar a=0\atop \bar c,\bar b=-1}\ .
\ee
Now the rational  anti--holomorphic parts of the two integrals assume either of the two forms $\tfrac{1}{\bar z(\bar z-\bar x)}$ and $\tfrac{1}{(\bar z-1)(\bar z-\bar x)}$, respectively,  and match those of the basis elements \req{Jmatrix}.
Performing similar manipulations in the holomorphic sector allows  expressing the complex integral \req{startI} in terms of the four basis elements \req{Jmatrix} and  to apply
\req{svRel}:
\begin{align}
\frac{1}{\pi}\Ic(x,\bar x)_{\bar a,\bar c=0\atop \bar b=-2}&=\fc{\hat a}{1-\hat b}\
\sv\int_0^xdw\ |w|^{\hat a+a}\ |1-w|^{\hat b+b}\ |w-x|^{\hat c+c}\nonumber\\
&+\fc{\hat a(1-\bar x)+\hat c}{(1-\bar x)(1-\hat b)}\ \sv\int_x^1dw\ |w|^{\hat a+a}\ |1-w|^{\hat b+b}\ |w-x|^{\hat c+c}\ .\label{Expressionsv}
\end{align}
The integrands in \req{Expressionsv} can be expanded w.r.t. small  $\hat a,\hat b,\hat c$. According to Appendix \ref{Appendixsv}, the $\sv$ map is then  applied on each period integral in the expansion.

For \req{Gcomplex} with $a=-1,b=1,c=0$ the two real integrals of \req{Expressionsv} are  related to $S_1(x)$ and $S_2(x)$ of Eqs.(\ref{s1def}) and (\ref{s2def}). More precisely, we define  the following two functions
\begin{align}
\hat S_1(x)&:=x(1-x)\ S_1(x)=\int_0^x\ dw\ w^{\hat a-1}\ (1-w)^{\hat b+1}\ (x-w)^{\hat c}\nonumber\\
&=x^{\hat a+\hat c}\ (1-x)^{2+\hat  b+\hat c}\ B(\hat a,1+\hat c)\ {}_2F_1\lf({2+\hat a+\hat b+\hat c,1+\hat c\atop 1+\hat a+\hat c};x\ri)\ ,\\
\hat S_2(x)&:=x(1-x)\ S_2(x)\ B(-\hat a,-\hat c)^{-1}=1+\hat b+\hat c x\ ,
\end{align}
which are related through the following identity:
\begin{align}\label{identy}
s(\hat a)\ \hat S_1(x)&-\pi\ \hat S_2(x)=s(\hat b)\int_x^1 w^{\hat a-1}\ (1-w)^{\hat b+1}\ (w-x)^{\hat c}\\
&=s(\hat b)\ x^{-\hat b}(1-x)^{2+\hat  b+\hat c}\ B(2+\hat b,1+\hat c)\ {}_2F_1\lf({2+\hat a+\hat b+\hat c,1+\hat c\atop 3+\hat b+\hat c};1-x\ri).\nonumber
\end{align}
The latter follows from the monodromy relation (9.80) of \cite{DiF} for $\tilde a=1+\hat b, \tilde b=-1+\hat a, \tilde c=-2$
$$s(\tilde b+\tilde c)\ \Ic_1(\tilde a,\tilde b,\tilde c;x)+s(\tilde c)\ \Ic_2(\tilde b,\tilde a,\tilde c;1-x)=s(\tilde a)\  \Ic_1(\tilde b,\tilde a,\tilde c;1-x)\ ,$$
 subject to $\tilde a+\tilde b+\tilde c=-2-\hat c$ with the integrals \req{Int1} and \req{Int2}.
We may apply\footnote{Note that, strictly speaking the $\sv$--map is only defined for period integrals. Hence, Eq.\req{putForm} may also be understood as defining equation for $\sv\; \hat S_2(x)$.} the single--valued map on \req{identy}:
\be\label{putForm}
\hat a\ \sv\; \hat S_1(x)- \sv\; \hat S_2(x)=\hat b\ \sv\int_x^1 w^{\hat a-1}\ (1-w)^{\hat b+1}\ (w-x)^{\hat c}\  ,
\ee
to cast \req{Expressionsv} into the form
\be
\frac{1}{\pi}\Ic(x,\bar x)=\fc{1}{(1-\bar x)}\fc{1}{\hat b(1-\hat b)}\lf\{\hat a\hat c\ \bar x\ \sv\;\hat S_1(x)-[\hat a(1-\bar x)+\hat c]\ \sv\;\hat S_2(x)\ri\}\ ,\label{ExpressionSV}
\ee
Eventually, with $\hat a=-i\lambda_4, \hat b=-i\lambda_2,\hat c=-i\lambda_3$ given in \req{abc} we can write \req{Gcomplex} as follows:
\begin{align}\label{Gcomplexsv}S_1(x)&\bar{I}_1(\bar{x}) +S_2(x)\bar{I}_2(\bar{x})
=  \fc{B(i\lambda_3,i\lambda_4) }{i\lambda_2}\
\frac{1}{x(1-x)}\ \fc{1}{1-\bar x}\nonumber\\
&\times \lf\{\ \lambda_3\lambda_4\ \bar x\ \sv\;\hat S_1(x)+i(\lambda_2+\bar x\lambda_4)\ \sv\;\hat S_2(x)\ \ri\}\ .
\end{align}

On the other hand, for \req{assign1} with $a=-1,b=2,c=-1$ the two real integrals of \req{Expressionsv} are related to $\tilde S_1(x)$ and $\tilde S_2(x)$ of Eqs.(\ref{s1tdef}) and (\ref{s2tdef}). More precisely, we define  the following two functions:
\begin{align}
\hat{S}_1'(x)&:=(1-x)\ \tilde S_1(x)=-\int_0^x\ dw\ w^{\hat a-1}\ (1-w)^{\hat b+2}\ (x-w)^{\hat c-1}\ ,\\
\hat{S}_2'(x)&:=(1-x)\ \tilde S_2(x)\ B(-\hat a,-\hat c)^{-1}=-[ 2+\hat b+(\hat c-1) x]\ ,
\end{align}
which are related through the same type of identity as \req{identy}:
\be\label{identy1}
s(\hat a)\ \hat{S}_1'(x)-\pi\ \hat{S}_2'
(x)=s(\hat b)\int_x^1 w^{\hat a-1}\ (1-w)^{\hat b+2}\ (w-x)^{\hat c-1}\ .
\ee
Consequently, we may repeat the steps from above to write \req{stcom} as follows:
\begin{align}\label{Gtcomplexsv}\tilde S_1(x)&\bar{I}_1(\bar{x}) +\tilde S_2(x)\bar{I}_2(\bar{x})
=  \fc{B(i\lambda_3,i\lambda_4) }{i\lambda_2}\
\frac{1}{(1-x)}\ \fc{1}{1-\bar x}\nonumber\\
&\times \lf\{\ \lambda_3\lambda_4\ \bar x\ \sv\;\hat S_1'(x)+i(\lambda_2+\bar x\lambda_4)\ \sv\;\hat S_2'(x)\ \ri\}\ .
\end{align}
In this way, we obtain the following sv integral  representation:
\begin{align}
G_{34}^{21}&(x,\bar{x})_{SV}=  \fc{B(i\lambda_3,i\lambda_4) }{i\lambda_2}\
\frac{1}{x(1-x)}\ \fc{1}{1-\bar x}\,\times \nonumber\\
&\left\{f^{a_1a_2b}f^{a_3a_4b}\big[\lambda_3\lambda_4\ \bar x\ \sv\;\hat S_1(x)+i(\lambda_2+\bar x\lambda_4)\ \sv\;\hat S_2(x)\big]
\right.
\nonumber \\[1mm]
& ~~+\left.xf^{a_1a_3b}f^{a_2a_4b}
\big[\lambda_3\lambda_4\ \bar x\ \sv\;\hat S_1'(x)+i(\lambda_2+\bar x\lambda_4)\ \sv\;\hat S_2'(x)\big]
\right\} .\label{repintsv}\end{align}

\section{Single-valued celestial amplitudes from inverted shadows}
Recall that the correlator (\ref{gcorf}), now written as a complex integral in Eq.(\ref{repint}), was constructed by a single-valued completion of the shadow transform of  a four-gluon celestial amplitude with respect to  gluon number 1. This prompts the immediate question  whether this integral can be interpreted as a shadow transform of certain ``single-valued'' celestial amplitude. In this section, we construct  such an amplitude, thus closing the loop and reaching ``another side'' of the starting point of part I.

At this point, it is convenient to return to the notation of part I, where the position of the shadow gluon operator was denoted by $z_1'$. Therefore, Eq.(\ref{xdef}) now reads
\be x=\frac{z_{1'2}z_{34}}{z_{1'3}z_{24}}\ ,\label{xdef1}\ee
while $z_1$ is reserved for the position of the original ``unshadowed'' gluon operator, and the cross-ratio
\be z=\frac{z_{12}z_{34}}{z_{13}z_{24}}\ .\label{zdef1}\ee
Up to this point, we have been considering the four-point shadow correlator
\begin{align}
\Big\langle\tilde\phi_{\tilde\D_1=1,+}^{a_1,-\epsilon}&(z_{1}',\zbar_{1}'
)\,
\phi_{\D_2,-}^{a_2,-\epsilon}(z_2,\bz_2)
\,\phi_{\D_3,+}^{a_3,+\epsilon}
(z_3,\bar z_3)
\,\phi_{\D_4,+}^{a_4,+\epsilon}(z_4,\bz_4)\Big\rangle\nonumber\\[1mm]
&=z_{1'2}^{1-i\lambda_2}\, z_{24}^{\, -1}\, z_{1'4}^{-1-i\lambda_4}\, z_{1'3}^{-2-i\lambda_3}\, \bz_{1'2}^{-1-i\lambda_2}\, \bz_{24}^{\, -1}\, \bz_{1'4}^{1-i\lambda_4}\, \bz_{1'3}^{-i\lambda_3}\, G^{21}_{34}(x,\bar{x})_{SV} \, , \label{eq:4pt1shadow}
\end{align}
where we used conformal invariance in order to recover dependence on individual points from the limit given in Eq.(\ref{gcorf}).

In order to rewrite the integrals (\ref{repint}) as shadow transforms with respect to $z_1$, we need to find a relation between the integration variable $w$ and $z_1$. As a hint, we note that in the original celestial amplitude written in Eq.(I.3.3) of part I, there was a relative factor of $z$ between the contributions associated with two color factors. A comparison with Eq.(\ref{repint}) indicates that
\begin{align}
x\frac{w-1}{w-x}&=z=\frac{z_{12}z_{34}}{z_{13}z_{24}}\ ,
\end{align}
therefore
\be
w=\frac{z_{14}z_{21'}}{z_{11'}z_{24}} .
\ee
Furthermore,
\be
w-1 = -\frac{z_{12}z_{1'4}}{z_{11'}z_{24}}\ , \quad w-x = \frac{z_{21'}z_{13}z_{1'4}}{z_{24}z_{11'}z_{1'3}}.\ee
We can substitute above relations into the integrals (\ref{repint}) and change the integration variables from $w$ to $z_1$. The corresponding Jacobian is $|{dw}/{dz_1} |^2$, with
\be
\frac{dw}{dz_1} = \frac{z_{1'2}z_{1'4}}{z_{11'}^2 z_{24}}.
\ee
In this way, we obtain
\begin{align}
\frac{1}{x(1-x)}&\int\!\! d^2w\, w^{-1-i\lambda_4} \bar w^{-i\lambda_4} (w-1)^{1-i\lambda_2} (\bar w-1)^{-2-i\lambda_2} (w-x)^{-i\lambda_3} (\bar w-\bar x)^{-i\lambda_3}\nonumber\\
&=\frac{z_{24}}{z_{23}z_{34}}z_{1'2}^{\,-1+i\lambda_2}\, z_{1'3}^{2+i\lambda_3}\, z_{1'4}^{1+i\lambda_4}\, \bz_{24}\, \bz_{1'2}^{1+i\lambda_2}\, \bz_{1'3}^{i\lambda_3}\, \bz_{1'4}^{\, -1+i\lambda_4} \nonumber\\[1mm]
&\qquad~~\times\! \int \frac{d^2 z_1}{z_{11'}^2}\, z_{12}^{1-i\lambda_2}\, z_{13}^{-i\lambda_3} \, z_{14}^{-1-i\lambda_4} \, \bz_{12}^{\, -i\lambda_2-2}\, \bz_{13}^{\, -i\lambda_3}\, \bz_{14}^{\, -i\lambda_4}\ ,
\end{align}
\begin{align}
-\frac{1}{1-x}&\int\!\! d^2w\, w^{-1-i\lambda_4} \bar w^{-i\lambda_4}(w-1)^{2-i\lambda_2} (\bar w-1)^{-2-i\lambda_2} (w-x)^{-1-i\lambda_3}(\bar w-\bar x)^{-i\lambda_3}\nonumber\\
&\quad=-\frac{1}{z_{23}} z_{1'2}^{\,-1+i\lambda_2}\, z_{1'3}^{2+i\lambda_3}\, z_{1'4}^{1+i\lambda_4}\, \bz_{24}\, \bz_{1'2}^{1+i\lambda_2}\, \bz_{1'3}^{i\lambda_3}\, \bz_{1'4}^{\, -1+i\lambda_4} \nonumber\\[1mm]
&\qquad~~~\times \int \frac{d^2 z_1}{z_{11'}^2}\, z_{12}^{2-i\lambda_2}\, z_{13}^{-1-i\lambda_3} \, z_{14}^{-1-i\lambda_4} \, \bz_{12}^{\, -i\lambda_2-2}\, \bz_{13}^{\, -i\lambda_3}\, \bz_{14}^{\, -i\lambda_4}  \ .
\end{align}
Now we see that indeed the correlator (\ref{eq:4pt1shadow}) has the form of a shadow transform:
\begin{align}
\Big\langle\tilde\phi_{\tilde\D_1=1,+}^{a_1,-\epsilon}&(z_{1}',\zbar_{1}'
)\,
\phi_{\D_2,-}^{a_2,-\epsilon}(z_2,\bz_2)
\,\phi_{\D_3,+}^{a_3,+\epsilon}
(z_3,\bar z_3)
\,\phi_{\D_4,+}^{a_4,+\epsilon}(z_4,\bz_4)\Big\rangle\nonumber\\[1mm]
&=\int \frac{d^2 z_1}{z_{11'}^2}
\Big\langle\phi_{\D_1=1,-}^{a_1,-\epsilon}(z_{1},\zbar_{1}
)\,
\phi_{\D_2,-}^{a_2,-\epsilon}(z_2,\bz_2)
\,\phi_{\D_3,+}^{a_3,+\epsilon}
(z_3,\bar z_3)
\,\phi_{\D_4,+}^{a_4,+\epsilon}(z_4,\bz_4)\Big\rangle_{\!SV}\ ,
\label{shtran}
\end{align}
with the celestial single-valued amplitude given by
\begin{align}
\Big\langle\phi_{\D_1=1,-}^{a_1,-\epsilon}&(z_{1},\zbar_{1}
)\,
\phi_{\D_2,-}^{a_2,-\epsilon}(z_2,\bz_2)
\,\phi_{\D_3,+}^{a_3,+\epsilon}
(z_3,\bar z_3)
\,\phi_{\D_4,+}^{a_4,+\epsilon}(z_4,\bz_4)\Big\rangle_{\!SV}\nonumber\\[1mm]
=&~~-\frac{1}{\pi}(1+i\lambda_2)B(i\lambda_3,i\lambda_4) z_{12}^{2-i\lambda_2}z_{13}^{-i\lambda_3} z_{14}^{-i\lambda_4} \bz_{12}^{\, -2-i\lambda_2} \, \bz_{13}^{\, -i\lambda_3} \bz_{14}^{\, -i\lambda_4} \nonumber\\[1mm] &~~\times\!\left\{f^{a_1a_2b}f^{a_3a_4b}\frac{1}{z_{12}z_{23}z_{34}z_{41}}
+f^{a_1a_3b}f^{a_2a_4b} \frac{1}{z_{13}z_{32}z_{24}z_{41}}\right\}.\label{eq:inv4pt}
\end{align}
Note that the above correlator is manifestly covariant under conformal transformations, with  correct weights of external gluon operators.
We end up with a rather surprising conclusion, that in the $\lambda_1=0$ soft limit, the single-valued MHV amplitude is given by the so-called Parke-Taylor (PT) denominators ``dressed'' by conformal factors. The denominator part is the same as in the MHV amplitude \cite{Parke:1986gb} and in Nair's superamplitude \cite{Nair:1988bq}.\footnote{This correlator has another interesting property. It satisfies the differential equation derived in Ref.\cite{Banerjee:2020vnt} for MHV gluon amplitudes in Mellin space (see also Ref.\cite{Hu:2021lrx}). This is rather unexpected because the single-valued amplitude (\ref{eq:inv4pt}) does not seem to be related to the original Mellin amplitude in any simple way.}
The corresponding CCFT correlator is given by
\begin{align}
{\cal G}_{34}^{21}(z,\bz)=&\lim_{z_1\to\infty,\bz_1\to\infty} \bz_1^2
\Big\langle\phi_{\D_1=1,-}^{a_1,-\epsilon}(z_{1},\zbar_{1}
)\,
\phi_{\D_2,-}^{a_2,-\epsilon}(1,1)
\,\phi_{\D_3,+}^{a_3,+\epsilon}
(z,\bar z)
\,\phi_{\D_4,+}^{a_4,+\epsilon}(0,0)\Big\rangle_{\!SV}\nonumber\\[1mm]
=&~\frac{(1+i\lambda_2)B(i\lambda_3,i\lambda_4)}{\pi z(1-z)} \big(f^{a_1a_2b}f^{a_3a_4b}-z
f^{a_1a_3b}f^{a_2a_4b} \big)\ .\label{eq:inv44pt}
\end{align}
{}From this point, we will proceed in a similar way as in the earlier part of the paper, by performing conformal block decompositions of the single-valued amplitude and exhibiting the factorization chains from four-point to two-point correlation functions.

In the $(12\rightleftharpoons 34)_{ \mathfrak{2}}$ channel, conformal blocks have the same form as in Eq.(\ref{blcs}), but now with
\begin{align}\nonumber&
h_{12}= -\textstyle \frac{i\lambda_2}{2}\ ,~  \bh_{12}=-\textstyle\frac{i\lambda_2}{2},\qquad\qquad
h_{34}=\textstyle\frac{i\lambda_3}{2}-\textstyle\frac{i\lambda_4}{2}
\ ,~  \bh_{34}=\textstyle\frac{i\lambda_3}{2}-\textstyle\frac{i\lambda_4}{2}\ ,\\[1mm] &
h_3+h_4=2+\textstyle\frac{i\lambda_3}{2}+\textstyle\frac{i\lambda_4}{2}=2-\textstyle
\frac{i\lambda_2}{2}\ , ~~~~~~~~\bh_3+\bh_4=\textstyle\frac{i\lambda_3}{2}+\textstyle\frac{i\lambda_4}{2}=-
\textstyle\frac{i\lambda_2}{2}\ ,
\end{align}
where we used $\lambda_2+\lambda_3+\lambda_4=0$. In order to expand the correlator (\ref{eq:inv4pt}) in this channel, we formally rewrite it, first as
\begin{align}
{\cal G}_{34}^{21}(z,\bz)
=\frac{1}{\pi}(1+i\lambda_2)& B(i\lambda_3,i\lambda_4)\bigg\{\frac{f^{a_1a_2b}f^{a_3a_4b}}{z(1-z)}
\, {}_2F_1\bigg({0,i\lambda_3\atop 1-i\lambda_2}; z\bigg)\, _2F_1\bigg({0,i\lambda_3\atop -i\lambda_2}; \bar{z}\bigg)\nonumber\\[1mm] &
-\frac{f^{a_1a_3b}f^{a_2a_4b}}{1-z}\, {}_2F_1\bigg({0,1+i\lambda_3\atop 2-i\lambda_2};z\bigg)\, _2F_1\bigg({0,i\lambda_3\atop -i\lambda_2};\bar{z}\bigg) \bigg\}\ ,\label{eq:inv4pt1}
\end{align}
and then proceed in the same way as in Section 3. As a result, we obtain
\be
{\cal G}_{34}^{21}(z,\bz)= \sum_{m=1}^\infty (s_m f^{a_1a_2b}f^{a_3a_4b} +\tilde{s}_m f^{a_1a_3b}f^{a_2a_4b})K^{21}_{34}\Big[ m-\frac{i\lambda_2}{2}, -\frac{i\lambda_2}{2}\Big](z,\bar{z}) \, ,
\ee
where
\begin{align}
s_m &= \frac{1}{\pi}(1+i\lambda_2)B(i\lambda_3,i\lambda_4) \frac{\Gamma(-i\lambda_2+m)\Gamma(i\lambda_4+m)}{\Gamma(1+i\lambda_4)\Gamma(-i\lambda_2+2m-1)} \, , \label{eq:Bm}\\
\tilde{s}_m &= -s_m -\frac{1}{\pi}(-1)^m(1+i\lambda_2)B(i\lambda_3,i\lambda_4) \frac{\Gamma(-i\lambda_2+m)\Gamma(i\lambda_3+m)}{\Gamma(1+i\lambda_3)\Gamma(-i\lambda_2+2m-1)} \, . \label{eq:Btm}
\end{align}
When compared with the block decomposition of the shadow correlator (\ref{gss1}), the present one contains only one subset of primaries, with $(h,\bar{h})  = (m-\frac{i\lambda_2}{2}, -\frac{i\lambda_2}{2})$. The spectrum starts at $m=1$ with a primary associated to a gluon with helicity $+1$ and dimension $\Delta=1-i\lambda_2=1+i\lambda_3+\lambda_4$. Its coefficient is
\begin{align}
s_1& f^{a_1a_2b}f^{a_3a_4b} +\tilde{s}_1 f^{a_1a_3b}f^{a_2a_4b}=\frac{1}{\pi}
(1+i\lambda_2)B(i\lambda_3,i\lambda_4)f^{a_1a_2b}f^{a_3a_4b}\nonumber\\
&=B(\Delta_3-1,\Delta_4-1)B(\Delta_2+1, 2-\Delta_2-\Delta_3-\Delta_4) \frac{f^{a_1a_2b}f^{a_3a_4b}}{\pi\Gamma(2-\Delta_2-\Delta_3-\Delta_4)} .\label{so1}
\end{align}
The corresponding operator appears at the leading order of the OPE of outgoing gluons \cite{Fan1903,Strominger1910}, as seen in Eq.(\ref{eq:OoutOout1}), which we repeat here for completeness:
\begin{align}
\phi_{\Delta_3,+}^{a_3,+\epsilon}(z_3,\bz_3)\phi_{\Delta_4,+}^{a_4,+\epsilon}(z_4,\bz_4) &\sim \frac{-i f^{a_3a_4x}}{z_{34}} B(\Delta_3-1,\Delta_4-1)\phi_{\Delta_3+\Delta_4-1,+}^{x,+\epsilon}(z_4,\bz_4)\, .\label{eq:OoutOout5}
\end{align}
Still at the leading order, we can fuse the resultant operator with the incoming gluon operator at $z_2$ \cite{Strominger1910}:
\begin{align}
\phi_{\Delta_2,-}^{a_2,-\epsilon}(z_2,&\bz_2) \phi_{\Delta_3+\Delta_4-1,+}^{x,+\epsilon}(z_4,\bz_4)\nonumber\\[1mm]
& \sim ~\frac{i f^{a_2xy}}{z_{24}}  \Big[
B(\Delta_2+1,2-\Delta_2-\Delta_3-\Delta_4)
\phi^{y,+\epsilon}_{\Delta_2+\Delta_3+\Delta_4-2,-}(z_4,\bz_4)
\nonumber\\[1mm]
& \qquad-B(\Delta_3+\Delta_4-2,2-\Delta_2-\Delta_3-\Delta_4)\phi^{y,-\epsilon}_{\Delta_2
+\Delta_3+\Delta_4-2,-}(z_4,\bz_4)\Big] \nonumber\\
& + ~\frac{i f^{a_2xy}}{\bz_{24}}  \Big[B(\Delta_2-1, 2-\Delta_2-\Delta_3-\Delta_4)\phi^{y,+\epsilon}_{\Delta_2+\Delta_3+\Delta_4-2,+}(z_4,\bz_4)
\nonumber\\[1mm]
& ~~~~~-B(\Delta_3+\Delta_4,2-\Delta_2-\Delta_3-\Delta_4)\phi^{y,-\epsilon}_{\Delta_2
+\Delta_3+\Delta_4-2,+}(z_4,\bz_4)\Big]\ . \label{eq:OinOout5}
\end{align}
This factorization chain ends with a two-point function of
$\phi_{\Delta_1=1,-}^{a_1,-\epsilon}(z_1,\bz_1)$ with the operators on the r.h.s.\ of the above OPE. By invoking the principle proposed in Section 4, that a non-vanishing two-point function must necessarily involve one incoming and one outgoing operator [see Eqs.(\ref{zero2pt},\ref{eq:2pt})], and conformal invariance, we conclude that the only non-vanishing contribution comes from the first operator on the r.h.s.\ of Eq.(\ref{eq:OinOout5}), $\phi^{y,+\epsilon}_{\Delta_2+\Delta_3+\Delta_4-2,-}(z_4,\bz_4)$. Indeed, with $\Delta_2+\Delta_3+\Delta_4-2=2-\Delta_1=1$,
\be
\Big\langle\phi_{\Delta_1,-}^{a_1,-\epsilon}(z_1,\bz_1)
\phi^{y,+\epsilon}_{2-\Delta_1,-}(z_4,\bz_4)
\Big\rangle = \frac{\delta^{a_1y}}{\pi\Gamma(\Delta_1-2) \, \bz_{14}^2} \qquad (\Delta_1=1).  \label{eq:o-o-2pt}
\ee
The normalization factor $[\pi\Gamma(\Delta_1-2)]^{-1}=[\pi\Gamma(2-\Delta_2-\Delta_3-\Delta_4)]^{-1}$
ensures that its product with the OPE coefficient of (\ref{eq:OoutOout5}) and the coefficient of the first term in (\ref{eq:OinOout5}) reproduces the conformal block coefficient (\ref{so1}). We conclude that the single-valued amplitude (\ref{eq:inv4pt}) and the corresponding CCFT correlator properly factorize in the $(12\rightleftharpoons 34)_{ \mathfrak{2}}$ channel, at least at the leading OPE order.

The analysis of other channels is very similar. In the $(14\rightleftharpoons 32)_{ \mathfrak{2}}$ channel, the conformal block expansion reads
\begin{align}
{\cal G}^{21}_{34}&(z,\bar{z}) = {\cal G}^{41}_{32}(1-z,1-\bar{z}) \nonumber\\
&=\sum_{m=1}^{\infty} (u_m f^{a_1a_2b}f^{a_3a_4b} +\tilde{u}_m f^{a_1a_3b}f^{a_2a_4b})K^{41}_{32}\Big[ m-1-\frac{i\lambda_4}{2},1-\frac{i\lambda_4}{2}\Big](1-z,1-\bar{z}) \ ,\label{uchan2}
\end{align}
where
\begin{align}
u_m &= \frac{1}{\pi}(1+i\lambda_2)B(i\lambda_3,i\lambda_4) \frac{\Gamma(-i\lambda_4+m-2)\Gamma(i\lambda_2+m-2)}{\Gamma(i\lambda_2-1)\Gamma(-i\lambda_4+2m-3)}\, ,\label{eq:Dm} \\
\tilde{u}_m &= \frac{1}{\pi}(1+i\lambda_2)B(i\lambda_3,i\lambda_4) (-1)^m \frac{\Gamma(-i\lambda_4+m-2)\Gamma(i\lambda_3+m)}{\Gamma(1+i\lambda_3)\Gamma(-i\lambda_4+2m-3)} \, . \label{eq:Dtm}
\end{align}
Here, the expansion starts at $m=1$ with a primary associated with a gluon with helicity $-1$ and dimension $\Delta=1-\lambda_4=
1+\lambda_2+\lambda_3$. The corresponding coefficient is
\begin{align}
&u_1 f^{a_1a_2b}f^{a_3a_4b} +\tilde{u}_1 f^{a_1a_3b}f^{a_2a_4b} = -f^{a_1a_4b}f^{a_2a_3b}\frac{1}{\pi}(1+i\lambda_2)B(i\lambda_3,i\lambda_4)
 \nonumber\\
&\qquad\qquad\qquad\qquad\qquad\qquad~~~~~= -f^{a_1a_4b}f^{a_2a_3b}\frac{1}{\pi}(1-i\lambda_4)B(i\lambda_3,i\lambda_4-1) \nonumber\\
&=- B(\Delta_3-1,1-\Delta_2-\Delta_3)B(\Delta_2+
\Delta_3,2-\Delta_2-\Delta_3-\Delta_4)
\frac{f^{a_1a_4b}f^{a_2a_3b} }{\pi\Gamma(2-\Delta_2-\Delta_3-\Delta_4)} \,  .\label{uu1}
\end{align}
In this case, we can start factorizing by taking the limit of $z_2\to z_3$, with the leading OPE
terms given in Eq.(\ref{eq:OinOout2}). This can be followed by $z_4\to z_2= z_3$,  which is described by a similar OPE. One ends up with sixteen two-point correlators involving $\phi_{\Delta_1=1,-}^{a_1,-\epsilon}(z_1,\bz_1)$ and one of the remaining operators, but
by following the same arguments as in Section 4, one finds that only one of them is non-vanishing. The product of the corresponding OPE and two-point coefficients does indeed yield the conformal block coefficient (\ref{uu1}).

The conformal block expansion in the  $(13\rightleftharpoons 42)_{ \mathfrak{2}}$ channel can be obtained in the same way, with the following result:
\begin{align}
 z^{2h_3}&\bz^{2\bh_3} {\cal G}^{21}_{34}(z,\bz)={\cal G}_{31}^{24}\Big(\frac{1}{z},\frac{1}{\bz}\Big)\nonumber\\
=& \sum_{m=1}^{\infty}(t_m f^{a_1a_2b}f^{a_3a_4b} +\tilde{t}_m f^{a_1a_3b}f^{a_2a_4b}) K^{24}_{31}\Big[ m-1-\frac{i\lambda_3}{2}, 1-\frac{i\lambda_3}{2}\Big] \Big( \frac{1}{z}, \frac{1}{\bar{z}}\Big),
\end{align}
where
\begin{align}
t_m &= -\tilde{t}_m -(-1)^m\frac{1}{\pi} (1+i\lambda_2)B(i\lambda_3,i\lambda_4) \frac{\Gamma(i\lambda_2-2+m)\Gamma(-i\lambda_3+m-2)}{\Gamma(i\lambda_2-1)\Gamma(-i\lambda_3+2m-3)} \, ,\label{eq:Fm} \\
\tilde{t}_m &= \frac{1}{\pi} (1+i\lambda_2)B(i\lambda_3,i\lambda_4)\frac{\Gamma(i\lambda_4+m)\Gamma(-i\lambda_3+m-2)}{
\Gamma(1+i\lambda_4)\Gamma(-i\lambda_3+2m-3)} \, . \label{eq:Ftm}
\end{align}
These conformal block spectrum and block coefficients match the $(14\rightleftharpoons 32)_{ \mathfrak{2}}$ channel expansion, Eqs.(\ref{uchan2})-(\ref{eq:Dtm}), upon exchanging $3\leftrightarrow 4$ and using Jacobi identity. Therefore, similar to $(14\rightleftharpoons 32)_{ \mathfrak{2}}$, at the leading OPE order, this channel has correct factorization properties.

To summarize, the single-valued amplitude (\ref{eq:inv4pt}) and the corresponding CCFT correlator (\ref{eq:inv44pt}) enjoy crossing symmetry and satisfy all bootstrap conditions, at least at the leading order of the OPE. In addition to helicity $\pm 1$ gluon operators with dimensions $\Delta=1+i\lambda$, the conformal block spectrum consists of primary field operators with
dimensions $\Delta=m+i\lambda$, with integer $m>1$. At each level,  two spin values, $J=m$ and $J=m-2$, are possible. For $m>1$, the primary fields appear in all group representations contained in the product of two adjoint representations. Extrapolating from the results of part I, we expect a larger, but always integer spin spectrum, to contribute to the celestial amplitude beyond the soft limit of $\lambda_1=0$.
\section{Summary and conclusions}
In the last section, we closed the loop and returned to where we started in part I -- to the four-gluon celestial amplitude.
Defined as the Mellin transform of  standard scattering amplitude, it is ill-defined as a CFT correlator because the positions of primary field operators are constrained by four-dimensional kinematics. Four-particle scattering events are planar, therefore the conformally invariant cross ratio of four complex coordinates is real. Such Mellin amplitude has no crossing symmetry and disagrees with the OPE of gluon operators. On the other hand, the single-valued amplitude derived in the last section is a ``good'' CFT correlator because it enjoys crossing symmetry and satisfies bootstrap self-consistency conditions in agreement with all known OPEs. Let us outline in more detail the path that led us to the end point so drastically different from the beginning of part I.

In part I, we extended the definition of Mellin amplitude to the entire complex plane by performing a shadow transformation on one of gluon operators. The conformal block decomposition of the shadow correlator revealed the presence of blocks with continuous complex spin. This was a sign of a much deeper problem. In its own way, the shadow correlator inherits the problems of the original Mellin amplitude, by being a multi-valued function of the cross ratio. In general, it is given by one of Appell's functions and has rather complicated monodromy properties, therefore it is difficult to use it as a starting point for constructing a single-valued correlator. Some significant simplifications occur, however, in the limit of the ``soft'' shadow operator with conformal dimension $\Delta=1$. Then all primary fields propagating in one of two-dimensional channels have the same antiholomorphic weight (although an infinite spectrum of holomorphic weights). The correlator reduces from Appell's to a hypergeometric function and bears a striking resemblance to the correlators of minimal models with null states at level 2. By following a procedure similar to minimal models, it is possible to construct a single-valued correlator by adding just one compensating function involving the shadow of the aniholomorphic block. In this way here, in part II, we obtained a single-valued completion of the shadow transform of celestial amplitude. Its conformal block spectrum consists of primary field operators with dimensions $\Delta=m+i \lambda$, with integer $m\geq 1$  and various, but always integer spin. Starting from $m=2$, the blocks appear in all group representations
contained in the product of two adjoint representations.

It is not surprising that the single-valued completion eliminates states with complex spin.
What is the most surprising, however, and somehow miraculous result of this work, is that the single-valued correlator is perfectly compatible with all known OPEs of gluon operators. It enjoys crossing symmetry and has correct factorization properties in all two-dimensional channels, {\em i.e}.\ it is fully ``bootstrapped.'' In the special case of $SU(2)$ gauge group, we also discussed  some non-leading OPEs, with two gluons fusing into $SU(2)$ singlets, triplets and quintuplets and with the respective OPE coefficients including Clebsch-Gordan coefficients.

The connection to minimal models facilitates one more step.
In the Coulomb gas formulation of minimal models, the correlators of primary fields (vertex operators) can be represented as complex integrals over the positions of ``charge-screening'' vertices. In minimal models with Verma modules degenerating at level 2, they can be represented by a single integral. We followed the approach of Dotsenko and Fateev and constructed such integral representation of the celestial correlator. We also made a connection to string theory,
where complex, closed string world-sheet integrals can be related to open string integrals via so-called single-valued projection. In this way, we constructed another integral representation of the celestial correlator, as a single-valued projection of a real-line integral. In this context, kinematic invariants (exponents of Koba-Nielsen factors) are tied to conformal dimensions of primary fields.

Recalling the beginning of  part I, the shadow transform was computed by integrating the celestial amplitude over the complex position of one of the gluon operators on a celestial sphere, with an appropriate integration factor determined by conformal weights. Now this shadow correlator has been completed to a single-valued function and further represented by a complex integral. By a suitable change of the integration variable, we brought it to a form of a shadow  transform of a function with the conformal transformation properties appropriate for a four-gluon celestial amplitude.
Undoing this shadow transform leads to a new correlation function, which we defined as the ``single-valued'' celestial amplitude. It has a form drastically different from the original ``Mellin'' celestial amplitude. It is defined over the entire complex plane and has correct crossing symmetry, OPE and bootstrap properties.

The single-valued amplitude is given by a simple expression. It consists of PT denominators ``dressed'' by the factors ensuring correct conformal transformation properties. The PT denominator part is the same as in the MHV amplitude  \cite{Parke:1986gb} and in Nair's superamplitude \cite{Nair:1988bq}. It is unlikely that the single-valued amplitude can be obtained by a simple ``improvement'' of the Mellin amplitude, in a more direct way than our ``shadowy'' detour. The superamplitude appears, however, in a natural way in Witten's formulation of string theory in twistor space \cite{Witten:2003nn}. This indicates that twistor space may be helpful in constructing celestial conformal field theory underlying single-valued amplitudes.

\vskip 1mm
\noindent {\bf Acknowledgments}\\[2mm]
This material is based in part upon work supported by the National Science Foundation
under Grant Number PHY--1913328.
Any opinions, findings, and conclusions or recommendations
expressed in this material are those of the authors and do not necessarily
reflect the views of the National Science Foundation.
Wei Fan is supported in part by the National Natural Science Foundation of China under Grant No.12105121.
 \newpage

\appendix
\section{SU(2) Clebsch-Gordan coefficients in vector basis}
Clebsch-Gordan coefficients are usually written in the angular momentum basis as
\be\langle J,M|\boldsymbol{\cdot}\big(|j_1,m_1\rangle\otimes|j_2,m_2\rangle\big)\equiv C_{J,M}^{j_1,m_1;j_2,m_2} \ee
In our case, gluons are in the adjoint (vector) representation of $SU(2)$, with $j_1=j_2=1$, therefore $J=0,1,2$.
On the other hand, in Feynman diagrams, gluon $SU(2)$ states $|a\rangle$ are usually labelled by vector indices $a=1,2,3$. The angular momentum and vector bases are related by a unitary transformation:
\be |a\rangle=\sum_{n=1}^3U_{an} |j=1,m=2-n\rangle.\ee
The matrix $U$  can be constructed by requiring that $J_3$ transforms from the diagonal matrix $J_3^{\,kl}=$ diag$(1,0,-1)$ to the vector rotation matrix $J_3^{\,ab}=-i\epsilon_{3}^{\,ab}$. It is given by
\begin{align}
U=
\left(
\begin{array}{ccc}
 \frac{i}{\sqrt{2}} & ~0 ~ & -\frac{i}{\sqrt{2}} \\
 \frac{1}{\sqrt{2}} &~ 0 ~  & \frac{1}{\sqrt{2}} \\
 0 & -i & 0 \\
\end{array}
\right).
\end{align}
In the vector basis, Clebsch-Gordan coefficients are given by
\be
C_{J,M}^{ab}\equiv\langle J,\,M|a,b\rangle
 = \sum_{k=1}^3\sum_{l=1}^3 U_{ak}U_{bl}C_{J,M}^{1,2-k ;1,2-l}\ .
 \ee
By using Clebsch-Gordan coefficients, listed for example in https://pdg.lbl.gov/2002/clebrpp.pdf, we obtain:
\begin{align}
C_{2,2}=
\left(
\begin{array}{ccc}
 -\frac{1}{2} & ~ \frac{i}{2} ~ & 0 \\
 \frac{i}{2} & ~ \frac{1}{2} ~ & 0 \\
 0 & ~0 ~ & 0 \\
\end{array}
\right),\qquad
C_{2,-2} =
\left(
\begin{array}{ccc}
 -\frac{1}{2} & ~-\frac{i}{2}~ & 0 \\
 -\frac{i}{2} & ~\frac{1}{2}~ & 0 \\
 0 & 0 & 0 \\
\end{array}
\right),
\end{align}

\begin{align}
C_{2,1}=
\left(
\begin{array}{ccc}
 0 & ~0~ & \frac{1}{2} \\
 0 & 0 & -\frac{i}{2} \\
 \frac{1}{2} & ~-\frac{i}{2}~ & 0 \\
\end{array}
\right),~~~~~\quad
C_{2,0} =
\left(
\begin{array}{ccc}
 \frac{1}{\sqrt{6}} & 0 & 0 \\
 0 & \frac{1}{\sqrt{6}} & 0 \\
 0 & 0 & -\sqrt{\frac{2}{3}} \\
\end{array}
\right),\quad
C_{2,-1}=
\left(
\begin{array}{ccc}
 0 & 0 & -\frac{1}{2} \\
 0 & 0 & -\frac{i}{2} \\
 -\frac{1}{2} & ~-\frac{i}{2}~ & 0 \\
\end{array}
\right),
\end{align}

\begin{align}
C_{1,1}=
\left(
\begin{array}{ccc}
 0 & 0 & \frac{1}{2} \\
 0 & 0 & -\frac{i}{2} \\
 -\frac{1}{2} & ~ \frac{i}{2}~ & 0 \\
\end{array}
\right),\quad
C_{1,0}=
\left(
\begin{array}{ccc}
 0 & ~ \frac{i}{\sqrt{2}}~ & 0 \\
 -\frac{i}{\sqrt{2}} & 0 & 0 \\
 0 & 0 & 0 \\
\end{array}
\right)
,\quad
C_{1,-1}=
\left(
\begin{array}{ccc}
 0 & 0 & \frac{1}{2} \\
 0 & 0 & \frac{i}{2} \\
 -\frac{1}{2} & ~-\frac{i}{2}~ & 0 \\
\end{array}
\right),
\end{align}
\begin{align}
C_{0,0}=
\left(
\begin{array}{ccc}
 \frac{1}{\sqrt{3}} & 0 & 0 \\
 0 & \frac{1}{\sqrt{3}} & 0 \\
 0 & 0 & \frac{1}{\sqrt{3}} \\
\end{array}
\right).~~~~~~~~~
\end{align}
Note that $C_{2,M}$ and $C_{0,M}$ are symmetric in vector indices $a,b$, while
$C_{1,M}$ is antisymmetric. Furthermore, $C_{J,M}^{*ab}=(-1)^{J-M}C_{J,-M}^{ab}$.
\section{The single--valued projection}\label{Appendixsv}
\def\sv{{\rm sv}}

In this appendix, we briefly review the single--valued map $\sv$ \cite{BrownGIA}.  Based on \cite{Stieberger:2013wea}, this map represents a direct relation between complex sphere and real iterated integrals \cite{Stieberger:2014hba}. For a review, we refer to \cite{Stieberger:2016xhs} and for a rigorous mathematical proof to \cite{Brown:2019wna}.
In particular, complex sphere integrals describing closed string amplitudes can be written as some projection of iterated real integrals representing open string amplitudes  \cite{Stieberger:2014hba}. The simplest example  arises for four--point scattering yielding the relation
\be\label{simple}
\frac{1}{\pi}\int_\IC d^2w\ \fc{|w|^{2 \hat a}\ |1-w|^{2 \hat b}}{|w|^2 (1-\bar w)\ }=
\sv\lf(\int_0^1 dx\ x^{ \hat a-1}\ (1-x)^{\hat b}\ri)\ ,
\ee
with parameters $\hat a,\hat b$ chosen such that both integrals converge.
While the integral on the l.h.s.~of \req{simple} describes a four--point closed string amplitude,
the integral on the r.h.s.~describes a four--point open string amplitude.
The projection $\sv$ is understood to act on the period integrals that arise after expanding both integrands w.r.t. small $\hat a,\hat b$, e.g. at the leading order in $\hat b$ we have:\footnote{Such relations also arise in the world--sheet sigma--model comparison of heterotic and type I open string gauge couplings \cite{Fan:2017uqy}.}
\be
\frac{1}{\pi}\int_\IC d^2w\ \fc{\ln|1-w|^2}{|w|^2(1-\bar w)}=\sv\int_0^1 dx\  \fc{\ln(1-x)}{x}=-\sv(\zeta_2)=0\ .
\ee
In fact, all periods appearing in \req{simple} are Riemann zeta functions
\be\label{Riemann}
\zeta_n=\sum_{k>0}k^{-n}\ ,\ n\geq 2\ ,
\ee
on which the  $\sv$--map acts as:
\be\label{ApplyMap}
\sv: \begin{cases}
\zeta_{2n+1}\mapsto 2\ \zeta_{2n+1},& n\geq1\ ,\\
\zeta_2\mapsto 0\ .&
\end{cases}
\ee
This map represents the single--valued projection $\sv$. It is called projection since, e.g.
$\zeta_2$--terms are
projected out. More generally, $\sv$ represents a morphism acting on the space of multiple zeta values (MZVs)
\be\label{mzetas}
\zeta_{n_1,\ldots,n_r}:=\zeta(n_1,\ldots,n_r)=
\sum\limits_{0<k_1<\ldots<k_r}\ \prod\limits_{l=1}^r k_l^{-n_l}\ \ \ ,\ \ \ n_l\in\IN^+\ ,\ n_r\geq2\ ,
\ee
mapping the latter to a subspace
of MZVs, namely the single--valued multiple zeta values (SVMZVs) \cite{BrownGIA}:
\be\label{trivial}
\SV(n_1,\ldots,n_r)\in\IR\ .
\ee
The numbers \req{trivial} can be obtained from the MZVs \req{mzetas} by generalizing the map \req{ApplyMap}\ to the full space of MZVs \cite{BrownGIA}:
\be\label{mapSV}
\sv: \z(n_1,\ldots,n_r)\mapsto\ \SV( n_1,\ldots,n_r )\ .
\ee
The map \req{mapSV}\ has been constructed\footnote{Strictly speaking, the map $\sv$ is defined in the Hopf algebra $\Hc$ of motivic MZVs $\zeta^m$.
In this algebra $\Hc$ the homomorphism $\sv: \Hc\ra\Hc^\sv$, with $\z^m(n_1,\ldots,n_r)\mapsto\SVM(n_1,\ldots,n_r)$ and
$\SVM(2)=0$ can be constructed, \cf \cite{BrownGIA} for details.} by Brown in Ref.\cite{BrownGIA}, where also SVMZVs have been studied from a mathematical
point of view.
For instance, we have $\SV(5,3)=\sv\;\mz{5,3}=-10 \mz{3} \mz{5}$ and $\SV(7,3)=\sv\;\mz{7,3}=-28\mz{3} \mz{7}-12\mz{5}^2$.

A generalization of \req{simple} is the relation
\be\label{Simple}
-\frac{\bz}{\pi}\int_\IC d^2w\ \fc{|w|^{2 \hat a}\ |1-w|^{2 \hat b}\ |w-z|^{2 \hat c}}{|w|^2 (\bar w-\bar z)\ }=
\sv\lf(\int_0^z dx\ x^{ \hat a-1}\ (1-x)^{\hat b}\ (z-x)^{\hat c}\ri)\ ,
\ee
with $z\in\IC$, which reduces to \req{simple} for $z\!=\!1$. Again, the complex integral can be computed by the single--valued map $\sv$ acting on the period integrals after expanding the integrand of the r.h.s. \req{Simple} w.r.t.
to small $\hat a,\hat b,\hat c$. This expansion gives rise to $\IQ$--linear combinations of MZVs \req{mzetas} and (Goncharov) multiple
polylogarithms (MPLs) $G$ depending on $z$:
\be\label{Goncharov}
G(a_1,a_2,\ldots,a_w;z)=\int_0^z\fc{dt}{t-a_1}\ G(a_2,\ldots,a_w;t)\ ,\ n_j\in\IN,\ n_r>1,\ a_j,z\in\IC ,
\ee
with $G(;z)=1$.
The single--valued projection on MPLs has  been constructed\footnote{In fact, SMVZs can also be defined as single--valued MPLs at argument $z\!=\!1$.} by Brown in \cite{Brown:2004ugm}, e.g.
\begin{align}
\sv\; G(a_1;z)&=G(a_1;z)+\ov{G(a_1;z)}\ ,\nonumber\\
\sv\; G(a_1,a_2;z)&=G(a_1,a_2;z)+G(a_1;z)\;\ov{G(a_2;z)}+\ov{G(a_2,a_1;z)}\ ,\ {\rm etc.}\ ,
\end{align}
with $a_i\in\{0,1\}$, i.e.
\be
\sv\; G(0;z)=\sv\ln z=\ln|z|^2\ \ \ ,\ \ \ \sv\; G(1;z)=\sv\ln (1-z)=\ln|1-z|^2\ ,
\ee
and
\be
\sv\; G(0,1;z)=G(0,1;z)+G(0;z)\ G(1;\bar z)+G(1,0;\bar z)\ ,\ {\rm etc.}\
\ee
Thus, at leading orders in $\hat b$ \req{Simple} yields
\begin{align}
-\frac{1}{\pi}\int_\IC d^2w\ \fc{\bar z\ \ln|1-w|^2}{|w|^2(\bar w-\bar z)}&=\sv\int_0^z dx\ \fc{\ln(1-x)}{x}=\sv\; G(0,1;z)\nonumber\\
&=-\Lc_2(z)+\Lc_2(\bar z)+\ln(1-\bar z)\ln|z|^2\ ,
\end{align}
with $G(0^{p-1},1;z)=-\Lc_p(z),\ p\geq1$.
To summarize,  the single--valued map $\sv$ projects the MZVs onto the space of SVMZVs (via \req{mapSV}) and MPs onto single--valued MPs. It can be applied separately for each factor in
the power series expansion w.r.t. small $\hat a,\hat b,\hat c$.

Generalizing \req{simple} leads to higher--point closed string amplitudes described by a matrix $J$ of  multi--dimensional complex integrations.
Then, the single--valued projection can be applied on a matrix $F$ of
Euler integrals comprising a  basis of real iterated integrals representing open string amplitudes and leading to the following matrix relation \cite{Stieberger:2014hba}:
\be\label{svRel}
J=\sv(F)\ .
\ee
Recently, the relation \req{svRel} has been extended to also include unintegrated points $z$
\cite{Vanhove:2018elu,Britto:2021prf}.
For the example \req{Simple}, this amounts to considering the following matrix comprising a  basis of real iterated integrals:
\be\label{Fmatrix}
F=\begin{pmatrix}
\ds{\int_0^z dw\ \fc{1}{w} }&\ds{\int_0^z\fc{1}{1-w} }\\[5mm]
\ds{\int_z^1 dw\ \fc{1}{x} }&\ds{\int_z^1\fc{1}{1-w} }
\end{pmatrix}
\times |w|^{ \hat a}\ |1-w|^{\hat b}\ |w-z|^{\hat c}\ .
\ee
On the other hand, the complex integral \req{Simple} is an element of the following basis
matrix:
\be\label{Jmatrix}
J=-\frac{1}{\pi}\begin{pmatrix}
\ds{\int_\IC d^2w\ \fc{1}{w}\fc{\bar z}{ \bar w(\bar w-\bar z)} }&\ds{\int_\IC d^2w\ \fc{1}{1-w}\fc{\bar z}{ \bar w(\bar w-\bar z)} }\\[5mm]
\ds{\int_\IC d^2w\ \fc{1}{w}\fc{1-\bar z}{ (\bar w-1)(\bar w-\bar z)} }&\ds{\int_\IC d^2w\ \fc{1}{1-w}\fc{1-\bar z}{ (\bar w-1)(\bar w-\bar z)} }
\end{pmatrix}
\times  |w|^{2 \hat a}\ |1-w|^{2 \hat b}\ |w-z|^{2 \hat c}\ .
\ee

\end{document}